\documentclass{article}
\usepackage{graphicx,epsfig}
\usepackage{graphicx,subfigure}
\usepackage{arxiv}
\usepackage{amsmath}
\usepackage[utf8]{inputenc} 
\usepackage[T1]{fontenc}    
\usepackage{hyperref}       
\usepackage{url}            
\usepackage{booktabs}       
\usepackage{amsfonts}       
\usepackage{nicefrac}       
\usepackage{microtype}      
\usepackage{lipsum}

\title{Effect of imposed shear on the dynamics of a contaminated two-layer film flow down a slippery incline}

\author{
  Muhammad Sani\\
  Department of Mathematics\\
   SRM Institute of Science and Technology\\
  Kattankulathur-603203, India\\
  \texttt{msdjunior9@gmail.com} \\
  \And
  Siluvai Antony Selvan\\
  Department of Mathematics\\
   SRM Institute of Science and Technology\\
  Kattankulathur-603203, India\\
  \texttt{antony.selvan22@gmail.com} \\
   \And
   Sukhendu Ghosh \\
  Department of Mathematics\\
   Indian Institute of Technology\\
  Jodhpur, Rajasthan-342037, India\\
  \texttt{sukhendu.math@gmail.com} \\
   \And
 Harekrushna Behera\thanks{Corresponding author}\\
  Department of Mathematics\\
  SRM Institute of Science and Technology\\
  Kattankulathur-603203, India\\
  \texttt{hkb.math@gmail.com} \\
}

\begin{document}
\maketitle

\begin{abstract}
The linear instability of a surfactant-laden two-layer falling film over an inclined slippery wall is analyzed under an influence of external shear which is imposed on the top surface of the flow. The free surface of the flow as well as the interface among the fluids are contaminated by insoluble surfactants. Dynamics of both the layers are governed by the Navier--Stokes equations, and the surfactant transport equation regulates the motion of the insoluble surfactants at the interface and free surface. 
Instability mechanisms are compared by imposing the external shear along and opposite to the flow direction. A coupled Orr--Sommerfeld system of equations for the considered problem is derived using the perturbation technique and normal mode analysis. The eigenmodes corresponding to the Orr--Sommerfeld eigenvalue problem are obtained by employing the spectral collocation method. The numerical results imply that the stronger external shear destabilizes the interface mode instability. However, a stabilizing impact of the external shear on the surface mode is noticed if the shear is imposed in the flow direction, which is in contrast to the role of imposed external shear on the surface mode for a surfactant laden single layer falling film. Further, in the presence of strong imposed shear, the overall stabilization of the surface mode by wall velocity slip for the stratified two-fluid flow is also contrary to that of the single fluid case. The interface mode behaves differently in the two zones at the moderate Reynolds numbers and higher external shear magnifies the interfacial instability in both the zones. An opposite trend is observed in the case of surface instability. Moreover, the impression of shear mode on the primary instability is analyzed in the high Reynolds number regime with sufficiently low inclination angle. Under such configuration, dominance of the shear mode over the surface mode is observed due to the weaker impact of the gravitation force on the surface instability. The shear mode can also be stabilized by applying the external shear in the counter direction of the streamwise flow. Conclusively, the extra imposed shear on the stratified two-layer falling film plays an active role to control the attitude of the instabilities.
\end{abstract}


\section{Introduction}
The applications of falling film are often encountered in the coating process, food manufacturing, condenser and biomedical engineering (\cite{anjalaiah2013thin}, \cite{anjalaiah2015effect}, \cite{zhao2019faraday}, \cite{aursand2019inclined} and \cite{lavalle2019suppression}). Several analysis were carried out to understand the hydrodynamic instability of homogeneous falling film (\cite{chin1986gravity}, \cite{cheng2000stability} and \cite{dandapat2008bifurcation}). Initially, the low Reynolds number falling film was analyzed at the small inclination angle using the perturbation technique  by \cite{yih1963stability}. It is observed that the small perturbations are amplified near the long-wave region while the disturbances are highly damped towards the short-wave region. Later, the asymptotic solution of falling film in both small and larger Reynolds number regions were developed by \cite{anshus1972asymptotic}. Further, the analysis was carried out at a high Reynolds number regime to show the impact of shear and surface modes on the primary instability of the falling film (\cite{chin1986gravity}). They observed that the surface mode stabilizes the fluid flow while the opposite trends are found in the case of shear mode.

Similarly there are some applications like liquid-liquid extraction, bolus dispersion process, earth's inner topography and atmosphere, where the physical problems can be modeled as the two-layer/multi-layer flows (\cite{jiang2004inertialess}, \cite{gao2008mechanism}, \cite{hu2008linear}, \cite{usha2013miscible} and \cite{govindarajan2014instabilities}). In the two-layer/multi-layer fluid system, the overall instabilities can be controlled by the interface between the fluids. The detailed analysis of interface-dominated viscosity and density stratified system for the free surface and the Couette-Poiseuille flow were analyzed by \cite{kao1965stability} and \cite{yih1967instability}, respectively. In the case of inertialess free surface flow with an uniform densities, the presence of high viscous fluid in the upper layer destabilizes the falling film (\cite{loewenherz1989effect}). This further results in the formation of wave at the surface and interface owing to the fluid-fluid and fluid-air interactions (\cite{chen1993wave}). For the plane Poiseuille flow, the presence of viscosity jump at the interface, and interaction between the fluid and end walls stabilize the shear mode due to the lubrication effect (\cite{hooper1989stability}). Such viscosity stratification can be used in transporting the high viscous fluid through pipes, in which the less viscous fluid reduces the frictional losses at the surface (\cite{joseph2013fundamentals}). This theory of linear stability analysis was later applied in the co-extrusion process in the polymer processing industry and validated with the experimental results by studying extensively the impact of density ratio, viscosity ratio, surface tension, and total flow rate on the instabilities of two-layer flows (\cite{vempati2010stability}). Further, \cite{usha2013miscible} studied the influence of viscosity stratification and miscibility on the instability of free falling film. A detailed overview on the instabilities of miscible viscosity stratified flows is provided by \cite{govindarajan2014instabilities}.

The primary instabilities play a crucial role in determining the qualities of many applications either by upgrading or degrading the end products (\cite{samanta2014effect}). Farther, numerous passive mechanisms like influence of wall velocity slip
(\cite{pascal1999linear,sadiq2008thin,samanta2011falling,ghosh2014channel,ghosh2016freesurface}) and insoluble surfactant (\cite{blyth2004effect, wei2005effect}) can be employed for controlling these instabilities in different flow systems. In particular, the slippery wall is often realized physically as an interface between the fluid and porous layer, which has profound environmental applications like the passage of water flow inside the crack and water-soil system (\cite{beavers1967boundary}). It was reported that the Darcian velocity in the porous layer rapidly changes to the interfacial velocity. \cite{pascal1999linear} studied the effect of wall slip on the linear stability of the homogeneous falling film and observed that the surface of film destabilizes while increasing the permeability of an inclined plane. Later, \cite{sadiq2008thin} investigated the homogeneous falling film down the porous inclined plane using the Benny's asymptotic expansion method, weakly-nonlinear and nonlinear analysis. Further, the depth-averaged model for the falling film down the slippery plane was developed by \cite{samanta2011falling} and observed that the presence of slip amplifies the nonlinear travelling waves. Moreover, \cite{ghosh2014channel} and \cite{ghosh2016freesurface} have shown the control of velocity slip on the instabilities due to continuous viscosity stratification.

In both the homogeneous and heterogeneous fluids, the dynamics of surface and interfacial tension can be understood by employing the insoluble surfactant. There are many studies in the literature analyzing an effect of insoluble surfactant on the homogeneous/heterogeneous falling films (\cite{gaver1990dynamics, grotberg1994pulmonary, blyth2004effect, anjalaiah2013thin, alhushaybari2020absolute} and \cite{bhat2020linear}). On adding the insoluble surfactant on a falling film, the surface tension is altered by surface concentration and it gives rise to the extra Marangoni mode (\cite{blyth2004effect}). It was observed that the existence of Marangoni mode suppresses the surface instabilities of the falling film. \cite{wei2005effect} investigated the marangoni destabilization of shear imposed surfactant laden falling film due to the shear mode at the small Reynolds number. A similar problem was also investigated by \cite{samanta2014shear} to understand the influence of imposed shear on a free falling film.  Other than this, the effect of insoluble surfactant on the falling film down the porous inclined plane is investigated by \cite{anjalaiah2013thin}. Besides, the presence of insoluble surfactant on the surface and interface of the two layer falling film is explored by \cite{samanta2014effect}. Recently, the work of \cite{wei2005effect} is extended to an arbitrary wavenumber region by \cite{bhat2019linear} and they observed that the surface mode stabilizes on applying the external shear opposite to the flow direction. Further, the effect of surfactant on the two-layer falling film down the inclined slipper plane has been investigated by \cite{bhat2020linear}.

In the aforementioned studies, the passive mechanisms like effects of insoluble surfactant and wall-slip parameter are used to control the instabilities at surface as well as interface on the single/two-layer falling films. However, additional instability exists, and occurring due to the presence of insoluble surfactant at the top surface and interface for the two-layer falling film down a slippery wall. 
Hence, there is a need of some active mechanism for controlling unwanted instabilities in such flows, which can be made possible by imposing the external shear stress in a constant rate on the top surface of the flow. In the present study, the effects of imposed shear and wall velocity slip (hydrophobic kind wall) are analyzed on controlling the surface and interfacial instabilities in the presence of insoluble surfactant. The Orr--Sommerfeld boundary-value problem for the proposed physical flow system is obtained using the perturbation technique and normal-mode analysis. The resultant eigenvalues are obtained by solving the system of equations using the spectral collocation method. The numerical analysis are widely carried out on the different flow parameters regimes for varying external shear rate to understand overall control mechanism of imposed shear on multiphase falling film.
The article is organized as follows: the mathematical formulation along with the base solution are discussed in Section \ref{MF}, the numerical results are discussed in Section \ref{RS} and the final outcomes of problem are highlighted in Section \ref{Conclusion}.

\section{Mathematical Formulation}  \label{MF}

\begin{figure}[h!]
\begin{center}
\includegraphics*[width=12cm]{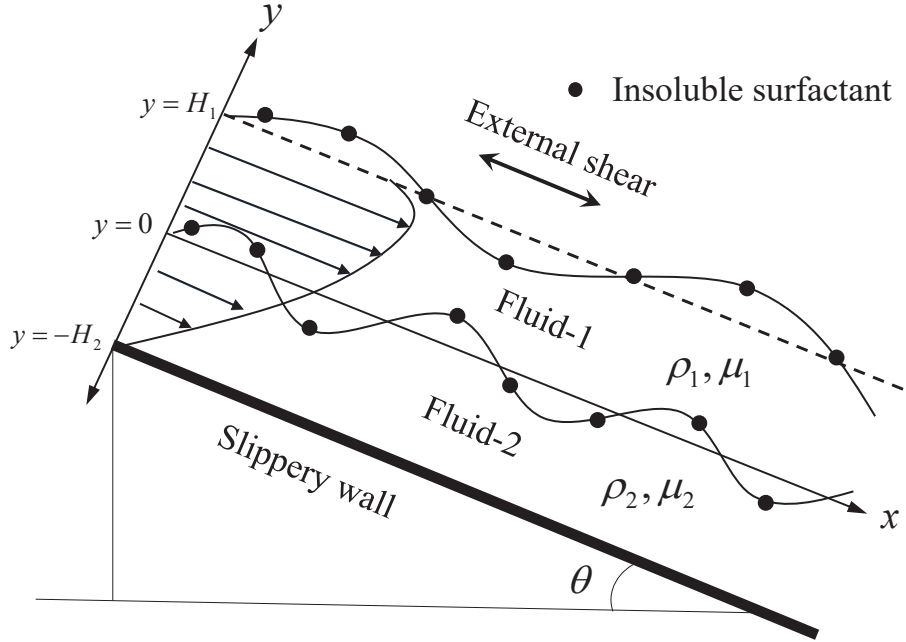}
\end{center}
	\caption{Schematic diagram for a two-layer thin film flow down in an inclined plane in the presence of insoluble surfactants.}\label{f1}
\end{figure}

A two-dimensional flow of two-layer immiscible and incompressible Newtonian fluids down an inclined slippery plane of angle $\theta$ is considered as in Fig.~\ref{f1}. The upper layer of thickness $H_1$ is known as fluid-1 having the density $\rho_1$ and the viscosity $\mu_1$, whereas the lower layer of thickness $H_2$ is denoted as fluid-2 having the density $\rho_2$ and the viscosity $\mu_2$. Further, the surface and interfacial elevations of each layers are represented as $\zeta^{(1)}$ and $\zeta^{(2)}$, respectively. The problem is studied in the Cartesian coordinate system with an origin located at the flat interface, where the $x$ and $y$-axes indicate the direction of streamwise and cross-streamwise flow for both the fluids. 
The motion of fluid-1 and fluid-2 is governed by the two-dimensional Navier--Stokes equations,
\begin{subequations}
\begin{eqnarray}
&u^{(j)}_x+v^{(j)}_y=0,
\\
&\rho_j~\big[u^{(j)}_t+u^{(j)}~u^{(j)}_x+v^{(j)}~u^{(j)}_y\big]=-p^{(j)}_x+\mu_j~(u^{(j)}_{xx}+u^{(j)}_{yy})+\rho_jg\sin\theta,
\\
&\rho_j~\big[v^{(j)}_t+u^{(j)}~v^{(j)}_x+v^{(j)}~v^{(j)}_y\big]=-p^{(j)}_y+\mu_j~(v^{(j)}_{xx}+v^{(j)}_{yy})-\rho_jg\cos\theta,
\end{eqnarray} \label{e1}
\end{subequations}
\noindent where $j=1$ and $j=2$ correspond the fluid-1 and fluid-2, respectively; $u^{(j)}$ and $v^{(j)}$ are known as the velocity of $j$\textsuperscript{th} layer fluid in streamwise and cross-streamwise directions; further, the pressure of $j$\textsuperscript{th} layer fluid and gravitational acceleration are represented as $p^{(j)}$ and $g$, respectively. In both the top surface and interface, the monolayered insoluble surfactant of concentration $\Gamma^{(j)}(x,t)$ diffuses into the each fluid resulting in the variations of local surface $\sigma^{(1)}$ and interfacial tension $\sigma^{(2)}$ of the fluid system. Thus, the relation between the surfactant concentration and the surface/interfacial tensions is expressed as
\begin{eqnarray}
& \sigma^{(j)}=\sigma^{(j)}_0-E_j(\Gamma^{(j)}-\Gamma^{(j)}_0),\label{e2}    
\end{eqnarray}
where $E_j$, $\sigma^{(j)}_0$ and $\Gamma^{(j)}_0$ are the surface elasticity, reference tension  and reference surfactant concentration, respectively. Moreover, $j = 1$ and $j = 2$ are respectively stand for the free surface and interface. For a static two-layer fluid, the kinematic conditions at the free surface and interface are expressed as
\begin{eqnarray}
v^{(j)}=\zeta^{(j)}_t+u^{(j)}~\zeta^{(j)}_x~~~~\mbox{at}~~~~y=\zeta^{(j)}(x,t).
\end{eqnarray}
Further, the dynamic boundary conditions corresponding to the tangential and normal shear stress balance at both the free surface and interface are given as
\begin{eqnarray}
&\mu_1\big[-4u^{(1)}_x\zeta^{(1)}_x+(u^{(1)}_y+v^{(1)}_x)~(1-(\zeta^{(1)}_x)^2)\big]=\big[\sigma^{(1)}_x\pm\tau\big]\,\sqrt{1+ (\zeta^{(1)}_x)^{2} } ~~\text{at} ~~ y = \zeta^{(1)}(x,t),\label{e4}\\
&p^{(1)} = \frac{2\mu_1}{\big[1+\big(\zeta^{(1)}_x\big)^2\big]}
\left\{u^{(1)}_x~(\zeta^{(1)}_x)^2-(u^{(1)}_y+v^{(1)}_x)~\zeta^{(1)}_x+v^{(1)}_y\right\}-\frac{\sigma_1~\zeta^{(1)}_{xx}}{\big[1+(\zeta^{(1)}_x)^2\big]^{3/2}}~~\text{at}~~y = \zeta^{(1)}(x,t),\label{e5}\\
&\mu_2\,\big[-4u^{(2)}_x\zeta^{(2)}_x+(u^{(2)}_y+v^{(2)}_x)(1-(\zeta^{(2)}_x)^2)\big]=\mu_1\,\big[-4u^{(1)}_x\zeta^{(2)}_x+(u^{(1)}_y+v^{(1)}_x)(1-(\zeta^{(2)}_x)^2)\big]\nonumber
\\ 
&\hspace{11.5cm}~~~\text{at}~~~~y=\zeta^{(2)}(x,t),\label{e6}
\end{eqnarray}
\begin{eqnarray}
&p^{(1)}+\frac{2\,\mu_1}{\big[1+(\zeta^{(1)}_x)^2\big]}\big[u^{(1)}_x\big(1-(\zeta^{(2)}_x)^2\big)+(u^{(1)}_y+v^{(1)}_x)~\zeta^{(2)}_x\big]+\frac{\sigma^{(2)}~\zeta^{(2)}_{xx} h_1}{\big[1+(\zeta^{(2)}_x)^2\big]^{3/2}}=p^{(2)}+\frac{2\mu_2}{\big[1+(\zeta^{(2)}_x)^2\big]}\nonumber\\
&\hspace{6.5cm}\big[u^{(2)}_x\big(1-(\zeta^{(2)}_x)^2\big)+(u^{(2)}_y+v^{(2)}_x)~\zeta^{(2)}_x\big]~~~\text{at}~~~~y=\zeta^{(2)}(x,t),\label{e7} 
\end{eqnarray}
where $\tau$ represents the external shear applied to the two-layer fluid. Further, the external shear applied along the direction of flow is denoted by $+\tau$, whereas $-\tau$
denotes the external shear applied opposite to the flow direction. At the interface, the horizontal and vertical velocity components of both the layers are continuous and result in the following boundary condition,
\begin{eqnarray}
u^{(1)}=u^{(2)}~~~~\mbox{and}~~~~v^{(1)}=v^{(2)}~~~~~\mbox{at}~~~~~y=\zeta^{(2)}(x,t).\label{e8}
\end{eqnarray}
Along the bottom of the two-layer flow, there is a velocity slip and thus the Navier--slip condition with no penetration of fluid resulting in,
\begin{eqnarray}
u^{(2)}=\beta u^{(2)}_y~~~~\mbox{and}~~~~v^{(2)}=0~~~~~\mbox{at}~~~~~y=-H_2,\label{e9}
\end{eqnarray}
where $\beta$ is the slip parameter. The transport equation governing the evolution of surfactant concentration at both the surface and interface are described as
\begin{eqnarray}
\Gamma^{(j)}_t+[(\textbf{u}^{(j)}.\textbf{t})\Gamma^{(j)}]_x+\Gamma^{(j)}\kappa~(\textbf{u}^{(j)}.\textbf{n})=\mathcal{D}_j\Gamma^{(j)}_{xx},\label{e10}
\end{eqnarray}
where $\textbf{u}^{(j)}.\textbf{t}$ and $\textbf{u}^{(j)}.\textbf{n}$ denote the velocities of the $j$\textsuperscript{th} layer fluid along the streamwise and cross-streamwise directions, respectively, and $\mathcal{D}_1$,  $\mathcal{D}_2$ are the surfactant diffusivity at the top surface and the interface, respectively. 

\subsection{Basic state solution and reference scales}

The governing equations and its associated boundary conditions of a steady state two-layer fluid flow are obtained by substituting $(u^{(j)},v^{(j)})=(U^{(j)}(y),0)$ and $p^{(j)}=P^{(j)}(y)$ in the above equations~\eqref{e1}--\eqref{e9}. By applying the local parallel flow assumptions, the solutions of steady state flow are given by
\begin{eqnarray}
&U^{(1)}(y)=\frac{\rho_1g\sin\theta H^2_1}{\mu_1}\left\{(1\pm\bar{\tau})\frac{y}{H_1}-\frac{y^2}{2H_1^2}+\frac{\delta(r\delta+2)}{2m}+\frac{\beta}{H_1}\frac{(\delta r+1)}{m}\pm\frac{\bar{\tau}}{m}\bigg(\frac{\beta}{H_1}+\delta\bigg)\right\},\label{e11}\\
&U^{(2)}(y)=\frac{\rho_1g\sin\theta H^2_1}{m\mu_1}\left\{(1\pm\bar{\tau})\frac{y}{H_1}-\frac{ry^2}{2H_1^2}+\frac{\delta(r\delta+2)}{2}+\frac{\beta}{H_1}(\delta r+1)\pm\bar{\tau}~\bigg(\frac{\beta}{H_1}+\delta\bigg)\right\},\label{e12}\\
&P^{(1)}(y)=\rho_1 g\cos\theta H_1~\bigg(1-\frac{y}{H_1}\bigg)~~~~\mbox{and}~~~~P^{(2)}(y)=\rho_1 g\cos\theta H_1~\bigg(1-\frac{ry}{H_1}\bigg),\label{e13}
\end{eqnarray}
where $\bar{\tau}=\tau/\rho_1H_1g\sin\theta$ is the non-dimensional form of the imposed shear stress. Further, the density ratio, depth ratio and viscosity ratio are expressed as $r=\rho_2/\rho_1$, $\delta=H_2/H_1$ and $m=\mu_2/\mu_1$, respectively. The characteristics velocity of two-layer film flow down the inclined slippery bottom in the presence of external shear stress is obtained by integrating the above base velocities over the film thicknesses of an each layer and averaging it with respect to the sum of thicknesses. This results in the following expression for the characteristic velocity
\begin{eqnarray}
U_c=\frac{\rho_1 g \sin\theta H^2_1}{\mathcal{K}\mu_1},\label{e14}
\end{eqnarray}
where
\begin{align*}
\frac{1}{\mathcal{K}}=\frac{1}{\delta+1}\left[\frac{1}{3}+\frac{\delta^2}{2m}+\frac{\delta}{m}+\frac{\delta^2r}{3m}+\frac{\delta^2r}{2m}+\frac{\bar{\tau}}{2}\bigg(1-\frac{\delta^2}{m}\bigg)\right]+\frac{\beta}{H_1}\bigg(\frac{\delta r}{m}+\frac{1}{m}\bigg)+\frac{\bar{\tau}}{m}\bigg(\frac{\beta}{H_1}+\delta\bigg).
\end{align*}
It is worth to note that the above expression recovers the characteristics velocity of two-layer falling film in the absence of slip bottom and external shear with no density stratification (\cite{kao1965stability}) by substituting $\tau=0$, $\beta=0$ and $r=1$. In the case of single layer fluid, the characteristics velocity of film flow having rigid bottom and no external shear (\cite{yih1963stability}) is obtained by substituting $r=1$, $\beta=0$, $m=1$, $\delta=0$ and $\tau=0$.
\begin{figure}[h!]
	\begin{center}
		\subfigure[$r=0.5$]{\includegraphics*[width=7cm]{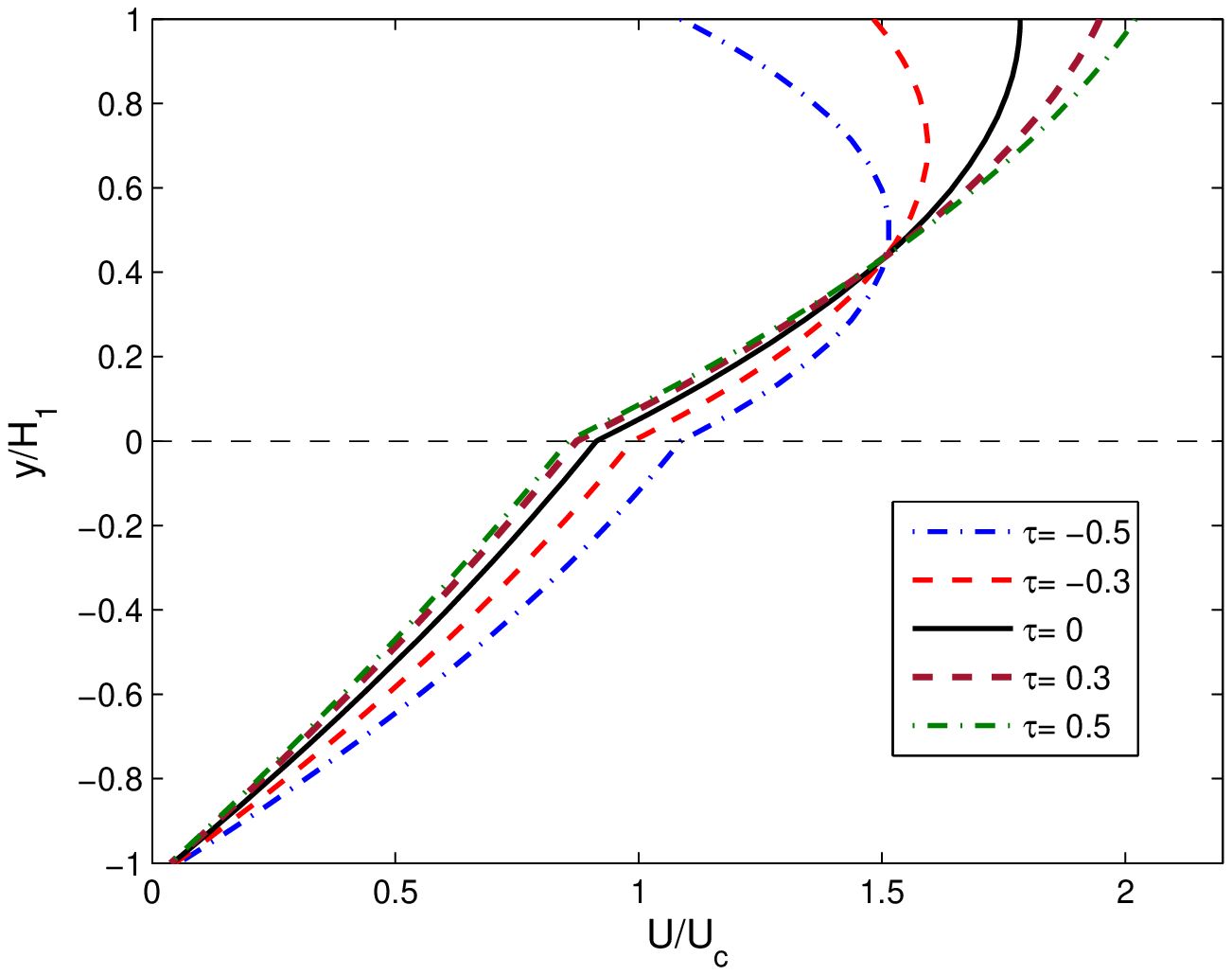}}
		\subfigure[$r=1$]{\includegraphics*[width=7cm]{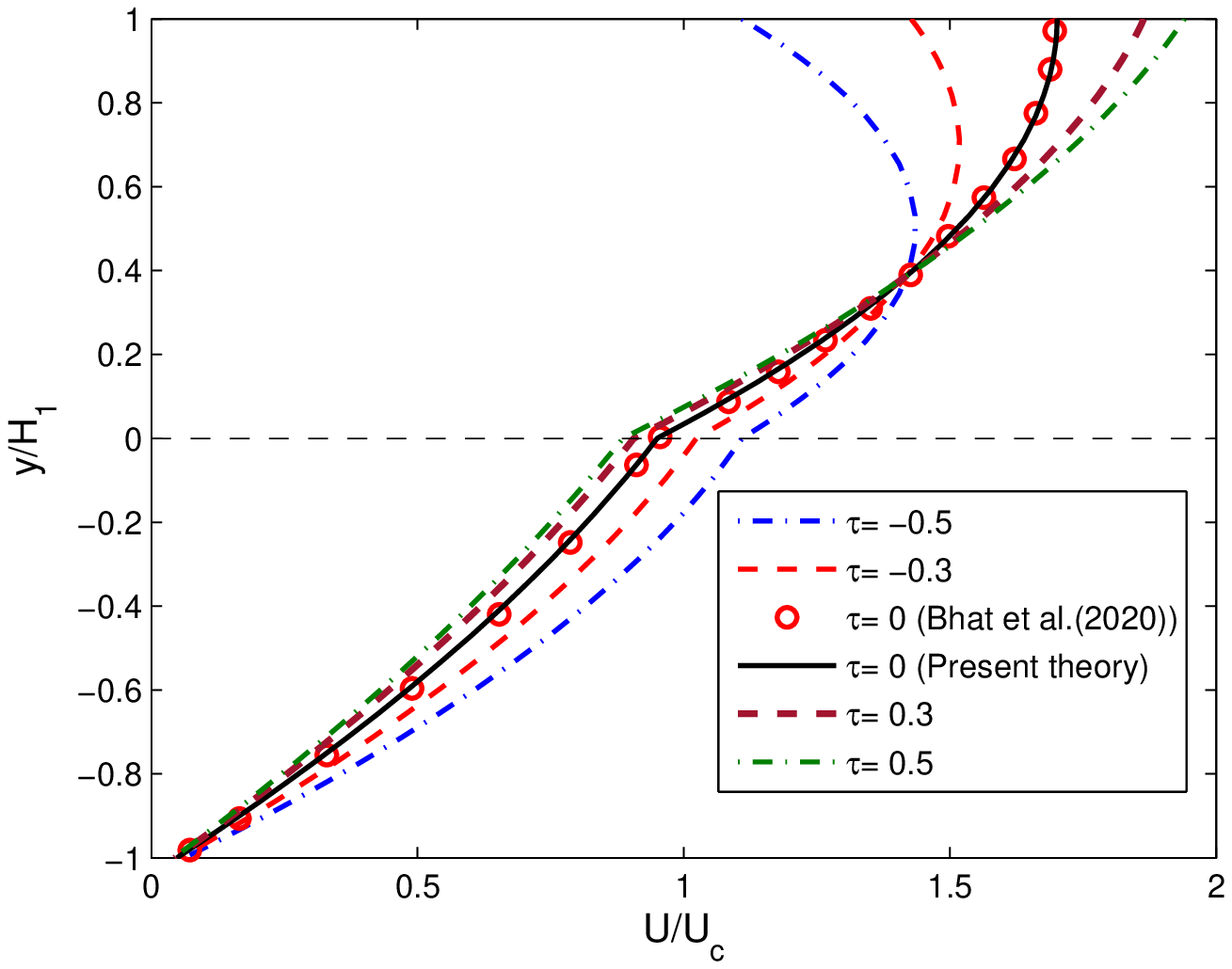}}
		\subfigure[$r=5$]{\includegraphics*[width=7cm]{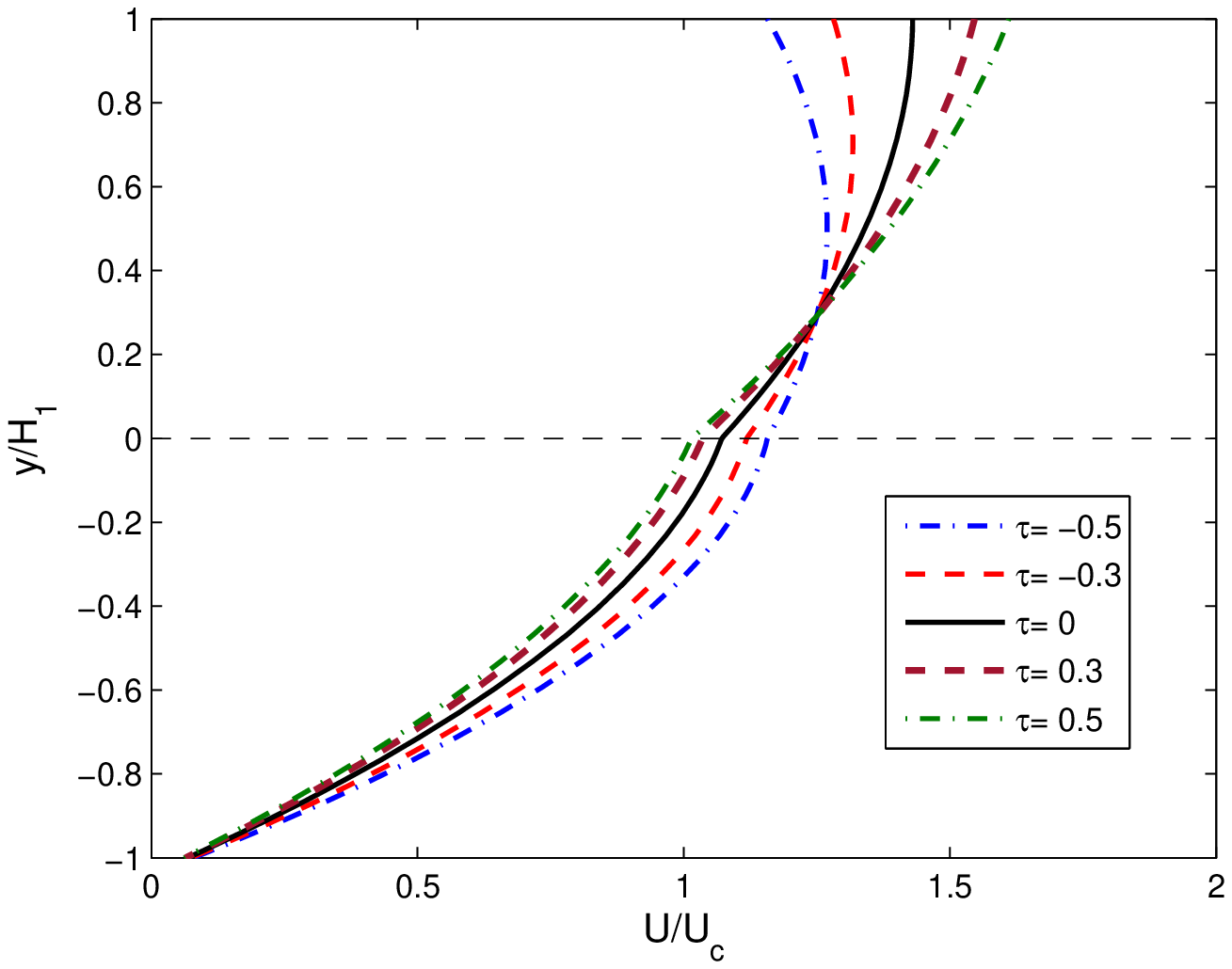}}
	\end{center}
	\caption{Variation of non-dimensional velocity of base state flow ($U/U_c$) with respect to the non-dimensional depth ($y/H_1$) for different values of non-dimensional external shear ($\tau$) showing the effect of density ratio $r$. The other constant flow parameters are $m=2.5$, $\delta=1$ and $\beta=0.04$.}\label{f2}
\end{figure}
Fig.~\ref{f2} exhibits the non-dimensional base state velocity profile as a function of non-dimensional depth for difference values of dimensionless imposed shear stress. The velocity profile in the upper layer advances on increasing the external shear stress from negative to positive values. The positive and negative $\tau$, restrictively, suggest the co and opposite direction of imposed shear with the flow direction. The present result retrieve the result of \cite{bhat2020linear} in the absence of external shear stress for $m=2.5$, $r=1$ and $\beta=0.04$ as in Fig.~\ref{f2}(b). When the shear stress is applied against the flow direction (i.e. for negative values of $\tau$), the net force ($\rho_1 g \sin\theta-\tau H^2_1$) acting  on the upper layer decreases and this weakens the base velocity along the surface in the upper layer fluid. Contrarily, for the shear stress along the flow direction (i.e. for positive values of $\tau$), the net force ($\rho_1 g \sin\theta+\tau H^2_1$) acting on the upper layer increases and enhances the base velocity of the upper layer fluid. On comparing the figures~\ref{f2}(a), (b) and (c), it is observed that the net force acting on the lower layer increases as compared to the upper layer as a result of increase in the value of $r$. 

In this study, the reference length, velocity and time scales are chosen as $H_1$, $U_c$ and $H_1/U_c$, respectively. Further, the reference scales for pressure, surfactant concentration and local surface tension for $j$\textsuperscript{th} layer fluid are assumed as $\rho_jU^2_c$, $\Gamma^{(j)}_0$ and $\sigma^{(j)}_0$, respectively. Using these scales, all the governing equations and its associated boundary conditions (Eqs.~\eqref{e1}--\eqref{e10}) are non-dimensionalized.

\subsection{Derivation of the OS eigenvalue problem}

The steady state flow is perturbed with infinitesimal disturbances 
such that the each dynamical variables are expressed as the sum of base state and perturbed state. These renewed variables are substituted in the dimensionless governing equations and boundary conditions, and linearized by removing the nonlinear terms with respect to the perturbations. The detailed procedure for obtaining the perturbed system of equations corresponding to a two-layer falling film is found in \cite{anjalaiah2013thin} and \cite{bhat2020linear}. Further, the perturbed variables are represented in the form of normal mode solution, which is given by $\hat{f}(x,y,t)=f(y)\,e^{\mathsf{i}k(x-ct)}$ where $\hat{f}$ being the arbitrary perturbed function with amplitude $f(y)$. Here, $k$ and $c$ are the wavenumber and complex phase speed, respectively. Expressing the linearized governing equation and its associated boundary condition in the form of normal mode amplitudes yield the Orr--Sommerfeld (OS) system of equation for the present physical model, which is given as follows:
\begin{subequations}
\begin{eqnarray}	
&(d_{yyyy}-2k^2d_{yy}+k^4)\,\phi^{(j)}-\mathsf{i}kRe_j\big[(U^{(j)}-c)~(d^2_y-k^2)~\phi^{(j)}-d_{yy}U^{(j)}\phi^{(j)}\big]=0,\label{e15a}\\
&\phi^{(j)}+(U^{(j)}-c)\eta^{(j)}=0~~~\mbox{at}~~y=1~~\mbox{for}~~j=1~~\mbox{and}~~y=0~~\mbox{for}~~j=2,\\
&d_y\phi^{(j)}+d_yU^{(j)}\eta_j+\bigg(U^{(j)}-c-\frac{\mathsf{i}k}{Pe_j}\bigg)\gamma^{(j)}=0~~~\mbox{at}~~y=1~~\mbox{for}~~j=1~~\mbox{and}~~y=0~~\mbox{for}~~j=2,\\
&(d_{yy}+k^2)\,\phi^{(1)}+d_{yy}U^{(1)}\eta_1+\frac{\mathsf{i}kMa_1}{Ca_1}\gamma^{(1)}=0~~~~\mbox{at}~~~~y=1,\\
&d_{yyy}\phi^{(1)}-3k^2d_y\phi^{(1)}-\mathsf{i}kRe_1\big[(U^{(1)}-c)d_y\phi^{(1)}-d_yU^{(1)}\phi_1\big]+\mathsf{i}k\bigg[2\mathsf{i}kU^{(1)}-\mathcal{K}\cot\theta-\frac{k^2}{Ca_1}\bigg]=0\nonumber\\
&\hspace{12.5cm}~~~~\mbox{at}~~~~y=1,\\
&d_y\phi^{(1)}-d_y\phi^{(2)}+(m-1)d_yU^{(2)}\eta^{(2)}=0~~~~\mbox{at}~~~~y=0,
\end{eqnarray}
\begin{eqnarray}
&\phi^{(1)}-\phi^{(2)}=0~~~~\mbox{at}~~~~y=0,\\
&\big[md_{yy}U^{(2)}-d_{yy}U^{(1)}\big]\eta^{(2)}+m(d_{yy}+k^2)\phi^{(2)}-(d_{yy}+k^2)\phi^{(1)}+\frac{\mathsf{i}kmMa_2}{Ca_2}\gamma^{(2)}=0~~~~\mbox{at}~~~~y=0,\\
&d_{yyy}\phi^{(1)}-3k^2d_y\phi^{(1)}-\mathsf{i}kRe_1\big[(U^{(1)}-c)d_y\phi^{(1)}-d_yU^{(1)}\phi_1\big]-m\biggl\{d_{yyy}\phi^{(2)}-3k^2d_y\phi^{(2)}-\mathsf{i}kRe_2\nonumber\\
&\hspace{2cm}\big[(U^{(2)}-c)d_y\phi^{(2)}-d_yU^{(2)}\phi_2\big]\biggr\}+\mathsf{i}k\bigg[\mathcal{K}(r-1)\cot\theta+\frac{mk^2}{Ca_2}\bigg]\eta^{(2)}=0~~~~\mbox{at}~~~~y=0,
\end{eqnarray}
\begin{eqnarray}
&d_y\phi^{(2)}-\beta d_{yy}\phi^{(2)}=0~~~~\mbox{at}~~~~y=-\delta\\
&\phi^{(2)}=0~~~~\mbox{at}~~~~y=-\delta,
\end{eqnarray}\label{e15}
\end{subequations}
where $Ma_j=E_j\Gamma^{(j)}/\sigma^{(j)}_0$, $Ca_j=U_c\mu_j/\sigma^{(j)}_0$, $Pe_j = U_c H_1/\mathcal{D}_j$, $Re_j=\rho_jU_cH_1/\nu_j$ are the Marangoni number, capillary number, Peclet number and Reynolds number of the $j$\textsuperscript{th} layer fluid, respectively. Note that the two Reynolds numbers are co-related by $Re_2 = (r/m)Re_1$.  Further, $d_y$ represents the derivative with respect to $y$, and $\phi^{(j)}$, $\eta^{(j)}$ and $\gamma^{(j)}$ denote the amplitudes of stream function, free surface/interface elevation and surfactant concentrations, respectively. There is no closed-form analytical solution for the above system of equations~\eqref{e15}(a)--(f) while solving for the arbitrary wavenumber and Reynolds number. Hence, the above system is solved numerically to obtain the eigenvalue $c$ corresponding to the OS system for the proposed physical problem. 

\subsection{Procedure for obtaining the eigenvalues}

In this subsection, the numerical methods used for obtaining the eigenvalues for the derived OS boundary value problem associated to the physical configuration is explained in detail. To carry out the linear stability analysis for arbitrary wavenumber and Reynolds number, the spectral collocation method (\cite{canuto2012spectral} and \cite{anjalaiah2013thin}) based on the Chebyshev polynomial is employed. The perturbation amplitudes of stream function in both the layers are written as the $N$-truncated series of Chebyshev polynomials, which is given as
\begin{eqnarray}
\phi^{(j)}=\sum_{n=0}^{N} \phi^{(j)}_n \mathcal{T}_n(y)~~~~\mbox{for}~~~~j=1,2,\label{e16}
\end{eqnarray}
provided $\mathcal{T}_n=\cos(n\cos^{-1}(y))~~\text{with}~~n=0,1,2,\dots N$, where $\mathcal{T}_n$ is the Chebyshev polynomial of first kind and Chebyshev points are used to discretize each layer over the domain $-1\leq y \leq 1$. Using the series expansion Eq.~\eqref{e16} in the OS system of equations~\eqref{e15}(a)--(f) yields the matrix eigenvalue problem, which can be written in the generalized form as
\begin{eqnarray}
\mathcal{A}\mathcal{X}=c\mathcal{B}\mathcal{X},\label{e17}
\end{eqnarray}
where the $\mathcal{A}$ and $\mathcal{B}$ are the square block matrices. Further, $c$ is the eigenvalue and its associated eigenvector is given by $\mathcal{X}=(\eta^{(1)},\eta^{(2)},\gamma^{(1)},\gamma^{(2)},\phi^{(1)}_1,\phi^{(1)}_2,\dots,\phi^{(1)}_N,\phi^{(2)}_1,\phi^{(2)}_2,\dots,\phi^{(2)}_N)$. The eigenvalues $(c = c_r + ic_i)$ are the complex phase speed for different perturbation waves, however, for studying the instability mechanism of the given physical problem, the primary dominant mode of disturbance is considered.

\section{Results and discussion} \label{RS}

In this Section the numerical results are discussed for a wide range of flow parameters. A Matlab 2014a subroutine is developed for solving the above eigenvalue problem (Eq.~\eqref{e17}) for obtaining the eigenmodes for various combinations of flow parameters. In general, in the case of two-layer flow in the presence of insoluble surfactant at the free surface and interface, there exist the two surfactant modes in addition to the unstable surface mode and interface mode (\cite{jiang2004inertialess, samanta2014effect} and \cite{bhat2020linear}). Results of the current study aims to characterize the control mechanism of the imposed external shear on these unstable modes. 

\begin{figure}[h!]
	\begin{center}
		\subfigure[]{\includegraphics*[width=7cm]{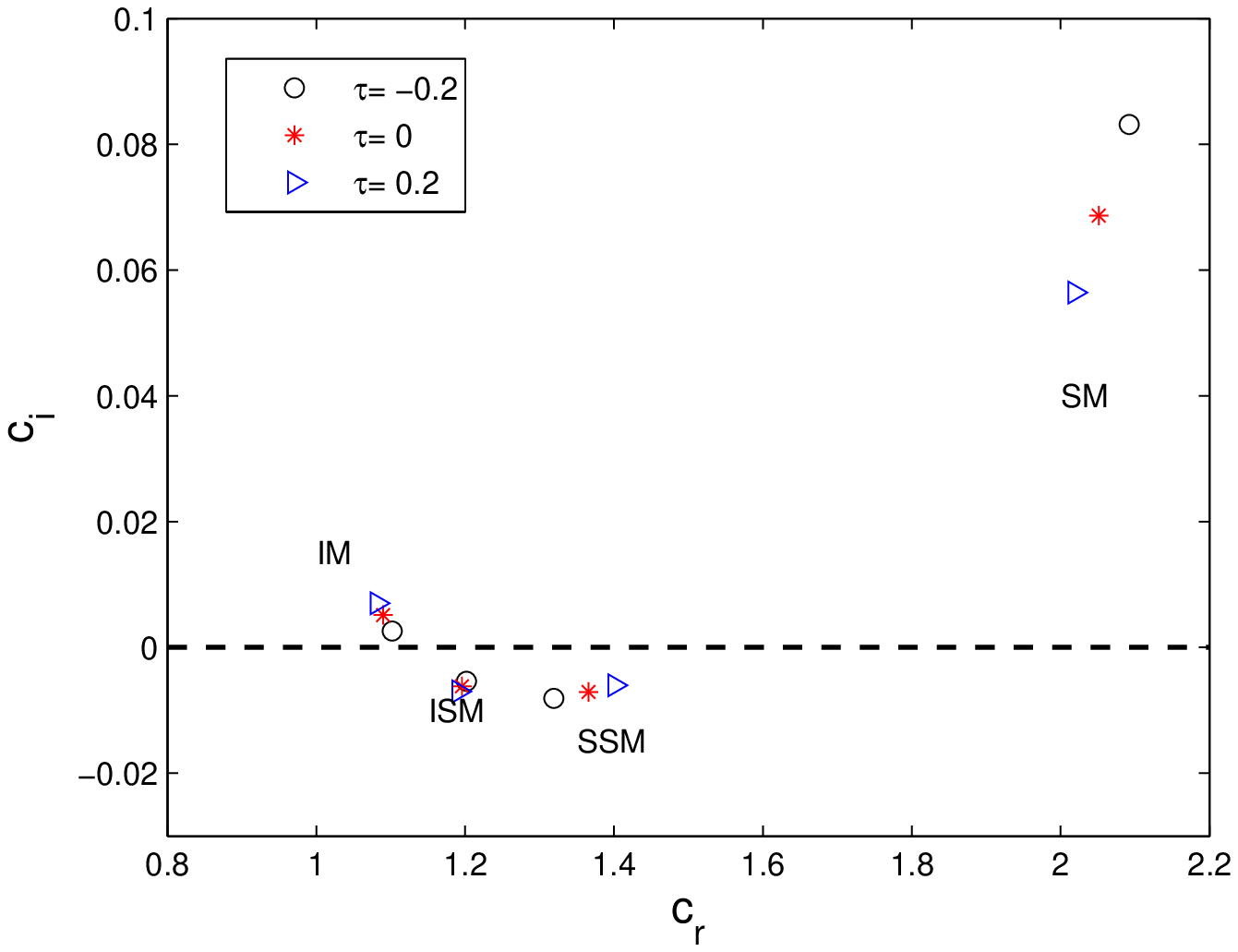}}
		\subfigure[]{\includegraphics*[width=7cm]{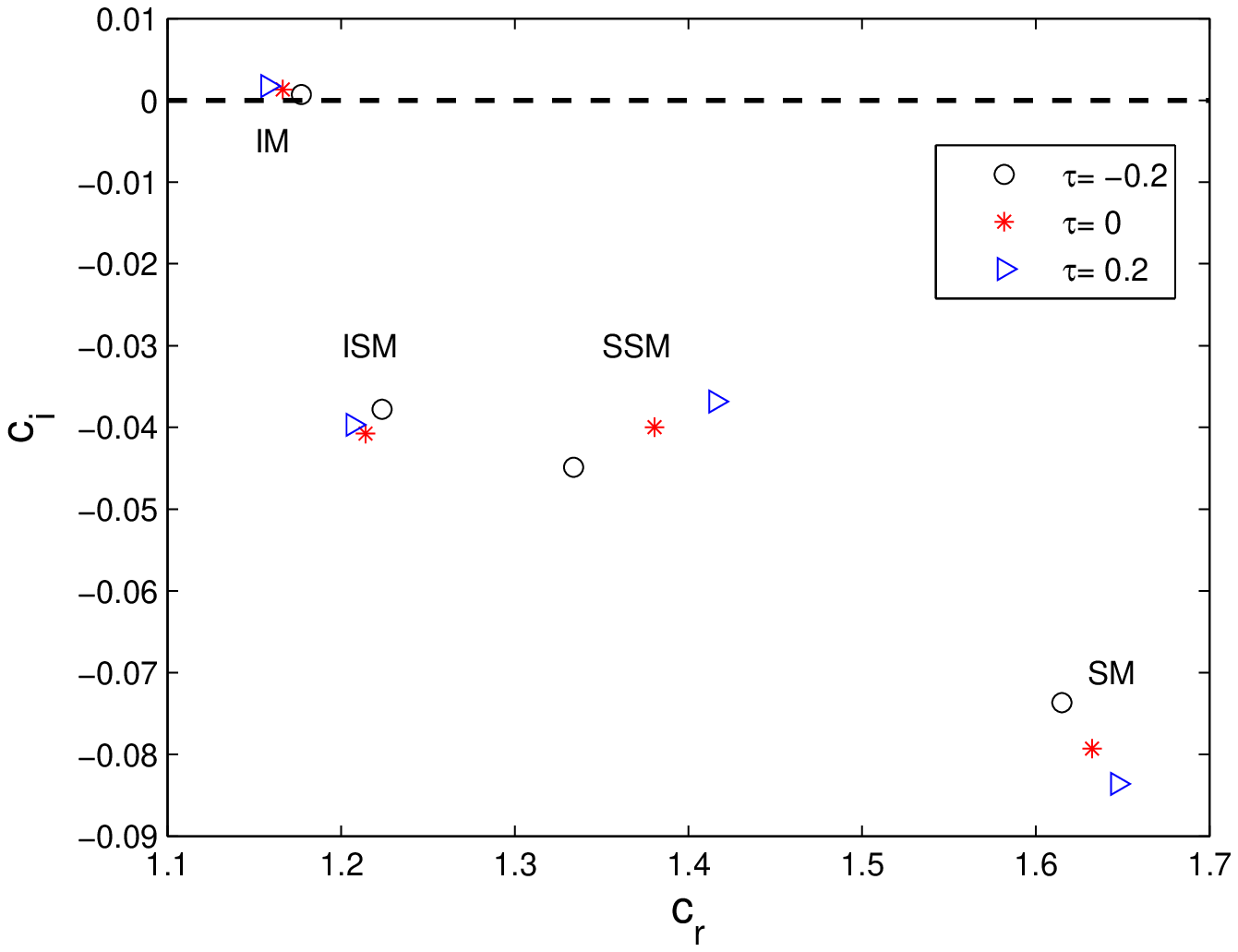}}
		\subfigure[]{\includegraphics*[width=7cm]{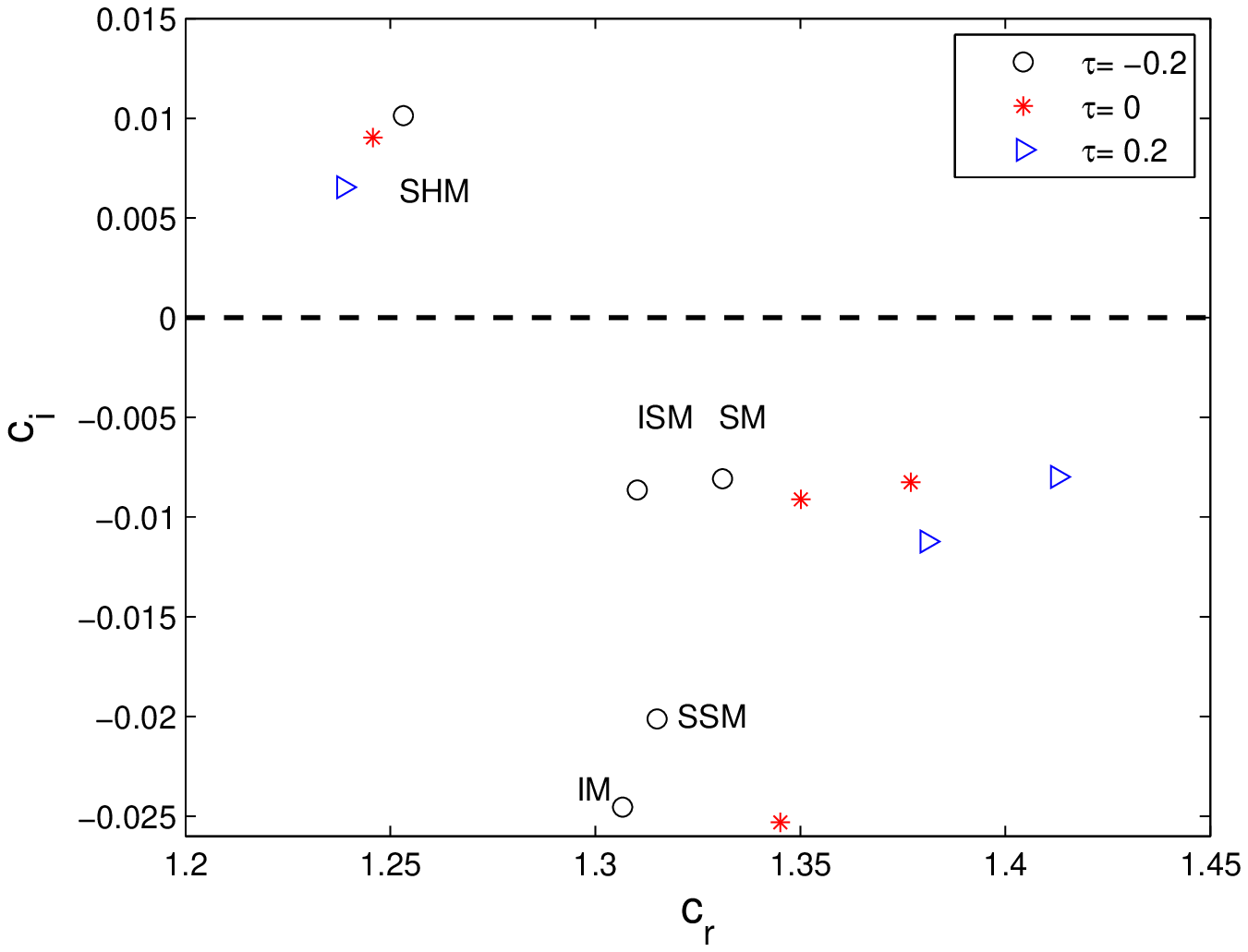}}
	\end{center}
	\caption{Distribution of eigenvalues showing the impact of external shear stress ($\tau$) on the destabilized (a) surface mode ($k=0.2$, $\theta=0.2$rad and $Re_1=5$), (b) interface mode ($k=1.5$, $\theta=0.2$rad and $Re_1=25$) and (c) shear mode ($k=0.2$, $\theta=0.5'$ and $Re_1=8000$). The other flow parameters are $m=0.5$, $r=1.1$, $Pe_1=Pe_2=10000$, $Ca_1=Ca_2=1$ and  $\beta=0.04$.}\label{f3}
\end{figure} 

In Figs.~\ref{f3}(a), (b) and (c), the effect of external shear on the destabilized surface mode, interface mode and shear mode is analyzed, respectively. Moreover, the modes corresponding to the free surface, interface, surface surfactant, interface surfactant and shear are denoted by SM, IM, SSM, ISM and SHM, respectively. It is evident from the Fig.~\ref{f3} that the eigenmodes satisfy the following condition based on the phase speed such as $c_r|_{SM}<c_r|_{SSM}<c_r|_{ISM}<c_r|_{IM}<c_r|_{SHM}$. In Figs.~\ref{f3}(a) and (b), the effect of external shear on the most destabilized surface and interface modes are analyzed at the moderate Reynolds number region, whereas the behaviour of external shear on the shear mode is analyzed at the high Reynolds number region is presented in Fig.~\ref{f3}(c). From Fig.~\ref{f3}(a), the stability of destabilized surface mode enhanced for increase in the external shear stress and it also restricts propagation of the surface mode by lowering the phase speed. This may be due to the decrease in net driving force on the surface for higher the external shear stress. In the case of destabilized interface mode as shown in Fig.~\ref{f3}(b), the external shear stress enhances the phase speed and further destabilizes the unstable interface mode. On comparing both the figures~\ref{f3}(a) and (b), the opposite trends are being observed under the influence of external shear in both the values of $c_r$ and $c_i$ as a result of momentum conservation. For the high Reynolds number and small inclination angle, the shear mode becomes unstable and dominant as shown in Fig.~\ref{f3}(c). On increasing the imposed shear rate, the shear mode stabilizes and the phase speed decreases. Thus, the shear instability abates for the stronger external shear.

\subsection{Influence of external shear on the interfacial mode (IM)}

Here, the effect of imposed shear on the interfacial mode instability, occurring due to the presence of a sharp interface between the two fluid layers, is analyzed for various flow configurations. The study carried out by \cite{bhat2020linear} in the absence of external shear suggests that there exist the two regions where the interfacial instability varies differently. Following the same, the effect of imposed shear on the interfacial mode is studied for two different zones, namely $mr<1$ and $mr>1$. 

\begin{figure}[h!]
	\begin{center}
		\subfigure[$mr<1$]{\includegraphics*[width=7cm]{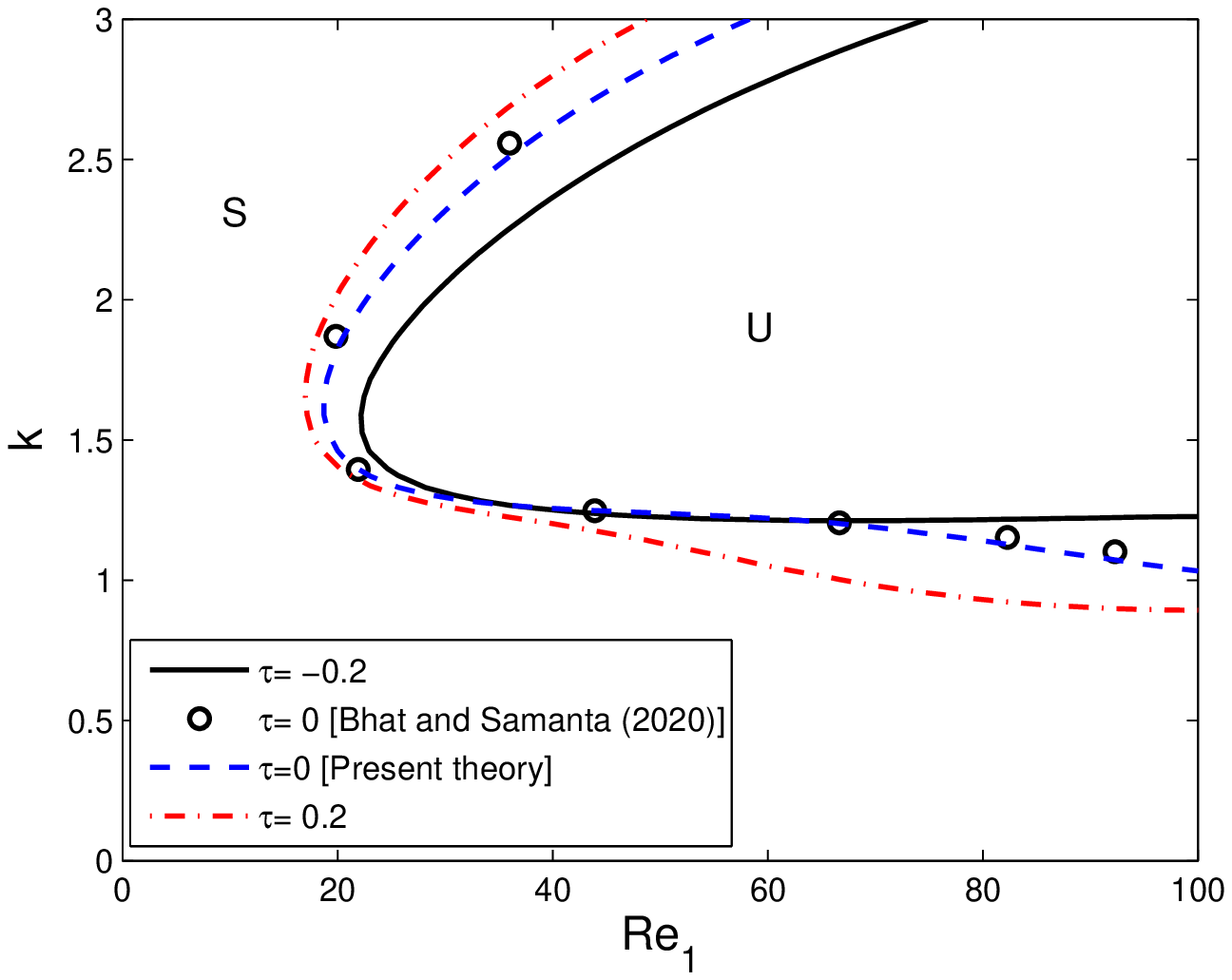}}
		\subfigure[$mr>1$]{\includegraphics*[width=7cm]{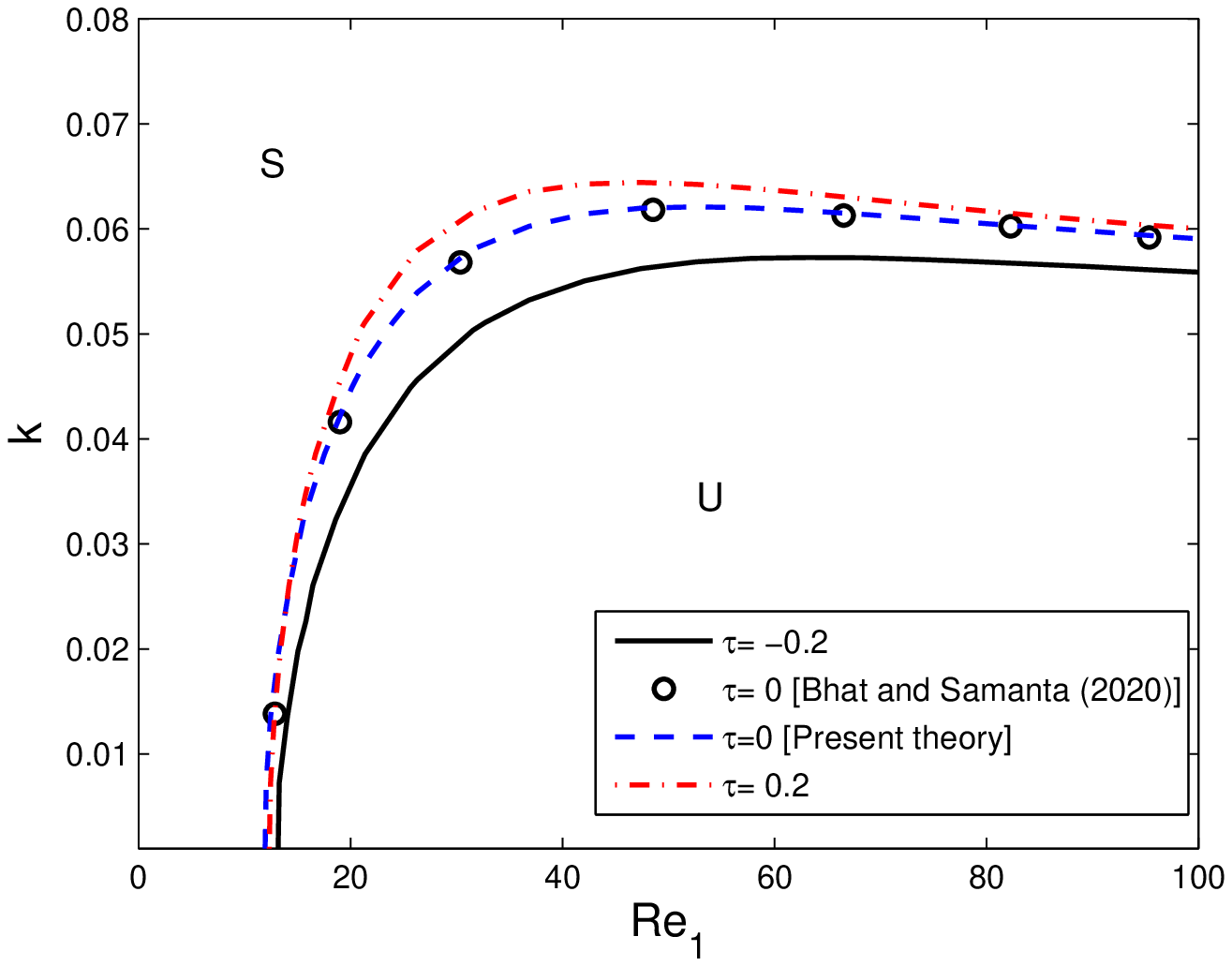}}
	\end{center}
	\caption{Marginal stability curve corresponding to varying external shear for (a) $mr<1$ with $r=1.1$ and (b) $mr>1$ with $r=2.4$. The other constant flow parameters are $m = 0.5$, $Pe_1=Pe_2=10000$, $Ca_1=Ca_2=1$, $\beta=0.04$, $\theta=0.2$ rad, $\delta=1$ and $Ma_1=Ma_2=0.1$.}\label{f4}
\end{figure}

In Figs.~\ref{f4}(a) and (b), the marginal stability curves showing the effect of external shear are plotted in $(Re_1,k)$ plane for $mr<1$ and $mr>1$, respectively. The results are well-matched with the existing theory of \cite{bhat2020linear} in the absence of external shear (i.e. $\tau=0$) for both $mr<1$ and $mr>1$. The stable and unstable regions are found on the either side of individual curves, which are marked as S and U, respectively. For $mr<1$ as in Fig.~\ref{f4}(a), there is a dominant unstable mode bandwidth occurs in the shorter-wave region, whereas the unstable mode bandwidth for $mr>1$ occurs near the long-wave region in Fig.~\ref{f4}(b). This is due to the effect of density stratification on the interface mode, which promotes the long-wave instabilities when the lower layer is more dense with a fixed viscosity ratio $m = 0.5$ and $mr>1$. 
Other than this, the presence of low viscous fluid in the lower layer further assist the interfacial instability as the velocity becomes faster at the interface. Moreover, the increasing external shear magnifies the unstable mode bandwidth for both $mr<1$ and $mr>1$. This may be due to the dissipation of surface energy by the imposed shear in the flow direction, which triggers the higher interfacial instability as a consequence of momentum conservation. However, to stabilize the interfacial instability one need to impose external shear $(\tau < 0)$ in the counter direction of the flow.

\begin{figure}[h!]
	\begin{center}
		\subfigure[$mr<1$]{\includegraphics*[width=7cm]{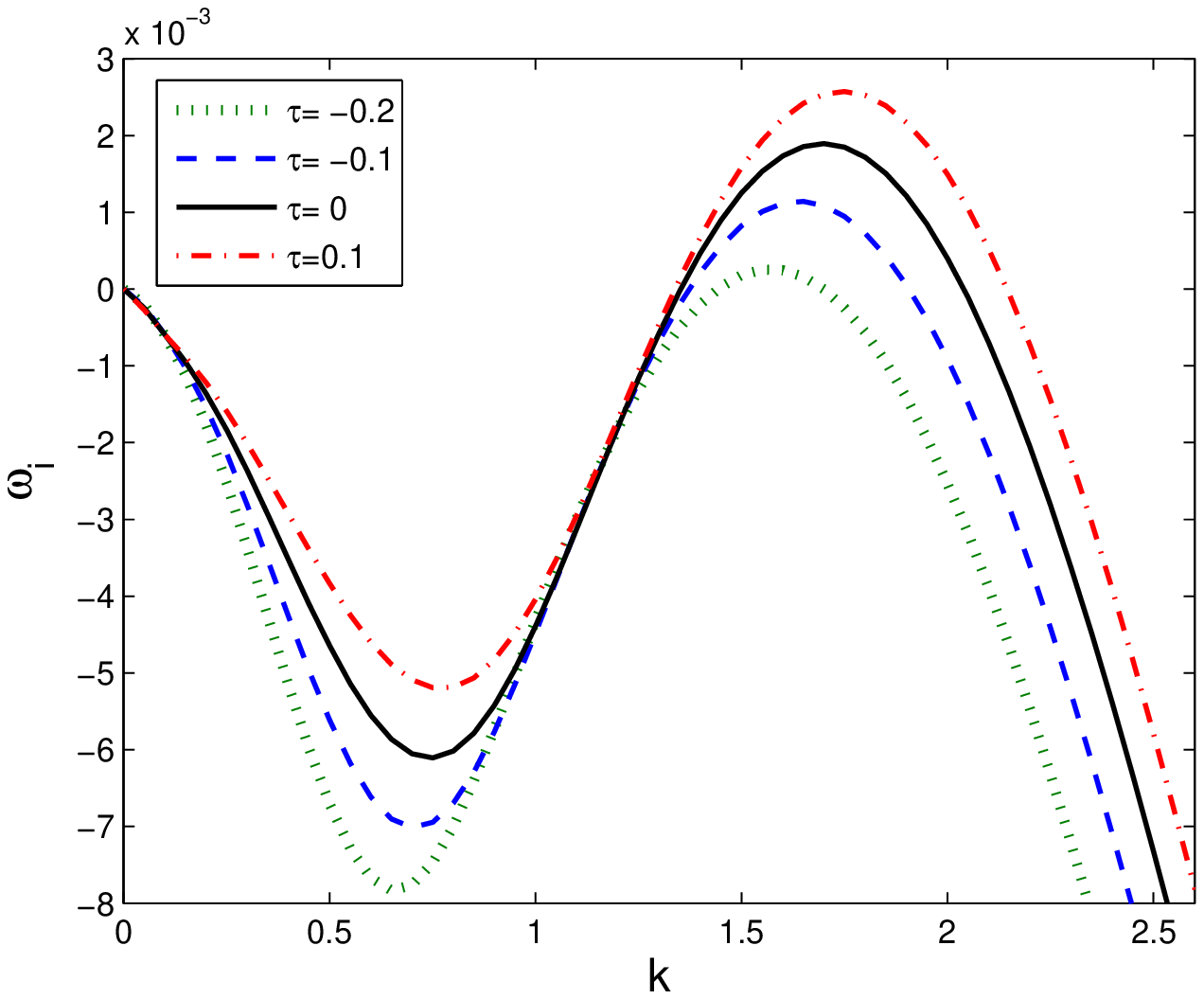}}
		\subfigure[$mr>1$]{\includegraphics*[width=7cm]{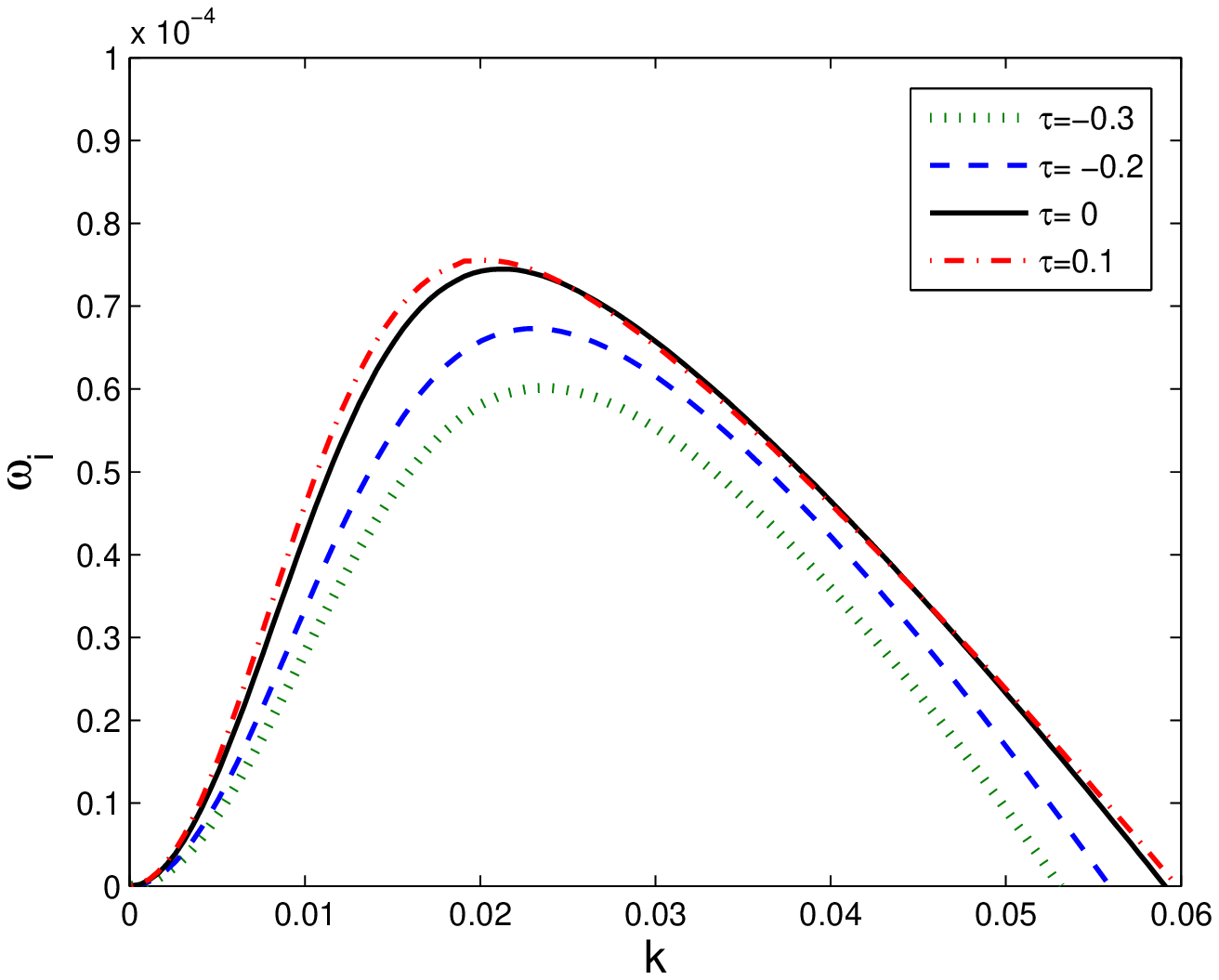}}
	\end{center}
	\caption{Scaled growth rate ($\omega_i$) as a function of wavenumber (k) showing the effect of external shear for (a) $mr<1$ ($r=1.1$, $Re_1=25$) and (b) $mr>1$ ($r=2.4$, $Re_1=100$). The other flow parameters are $m=0.5$, $Pe_1=Pe_2=10000$, $Ca_1=Ca_2=1$, $\beta=0.04$, $\theta=0.2$ rad, $\delta=1$ and $Ma_1=Ma_2=0.1$.}\label{f5}
\end{figure}
\begin{figure}[h!]
	\begin{center}
		\subfigure[$mr<1$]{\includegraphics*[width=7cm]{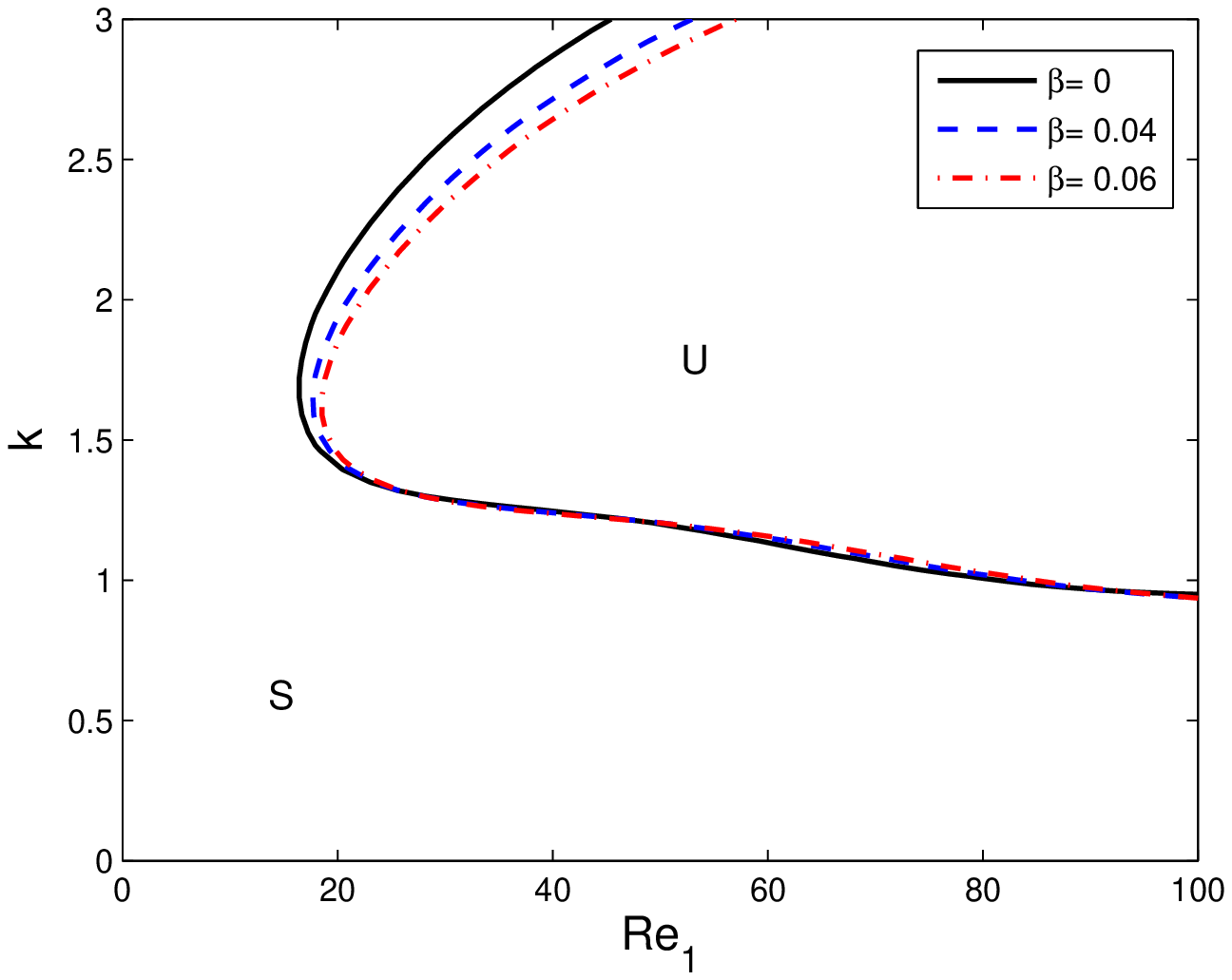}}
		\subfigure[$mr>1$]{\includegraphics*[width=7cm]{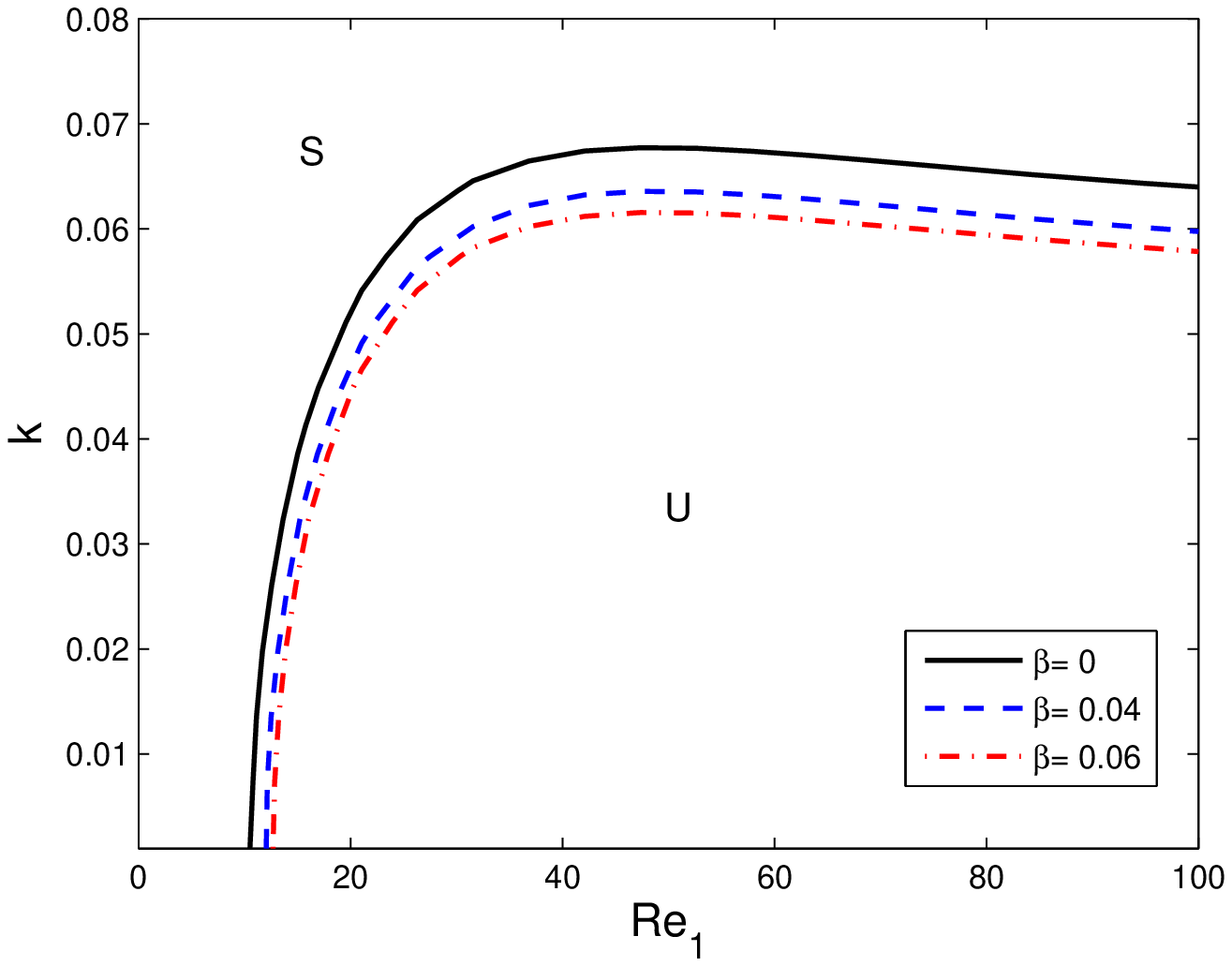}}
	\end{center}
	\caption{Marginal stability curve corresponding to varying slip parameter for (a) $mr<1$ with $r=1.1$ and (b) $mr>1$ with $r=2.4$. The other constant flow parameters are $m=0.5$, $Pe_1=Pe_2=10000$, $Ca_1=Ca_2=1$, $\beta=0.04$, $\theta=0.2$ rad, $\delta=1$ and $Ma_1=Ma_2=0.1$.}\label{f6}
\end{figure}

The scaled growth rate $(\omega_ = kc_i)$ corresponding to the unstable interface mode for $mr<1$ and $mr>1$ are respectively plotted in Figs.~\ref{f5}(a) and (b) for different values of $\tau$ at a particular $Re_1$. The growth rate attains negative and positive values for the long and moderate waves, respectively. This is evident from the  Fig.~\ref{f5}(a), where the shorter-wave instabilities are are prominent in the case of $mr<1$. On the other hand, the growth rate curves in Fig.~\ref{f5}(b) curve shows positive growth rate in the long-wave region. This occurs due to the presence of high dense fluid at lower layer with $mr>1$, which exposes to suppress the interfacial instability for larger wavenumber range. With increase in the values of external shear in Fig.~\ref{f5}(a), the instability enhances as a consequence of increasing net force acting on the upper layer. When the external shear is applied along the flow direction (i.e. for positive values of $\tau$ ), the increase in a net driving force near the interface is assisted by the external shear. While applying the external shear opposite to the flow direction (i.e. for negative values of $\tau$ ), the net driving force acting at the interface decreases. Thus, increasing the external shear value from negative to positive increases the interfacial instability and the same observation is made in the case of region $mr>1$.

It is evident from the Figs.~\ref{f4} and \ref{f5} that the positive value of $\tau$ has destabilizing effect on the interfacial flow, whereas the negative value shows the stabilizing effect on the interfacial mode. In Figs.~\ref{f6}(a) and (b), the effect of slip parameter on the marginal stability curve is plotted in the $(Re_1,k)$ plane for $mr<1$ and $mr>1$, respectively. The patterns of neutral curve is same as observed in the Fig.~\ref{f4}. However,  the presence of wall slip stabilizes the interfacial instability. On increasing the slip length, the velocity of the lower layer increases near the vivinity of the bottom wall. It results in a higher wall shear rate and redaction in frictional forces. Moreover, higher wall shear rate is well balanced by the interfacial shear and this scenario is more significant for $m<1$.

\begin{figure}[h!]
	\begin{center}
		\subfigure[$mr>1$]{\includegraphics*[width=7cm]{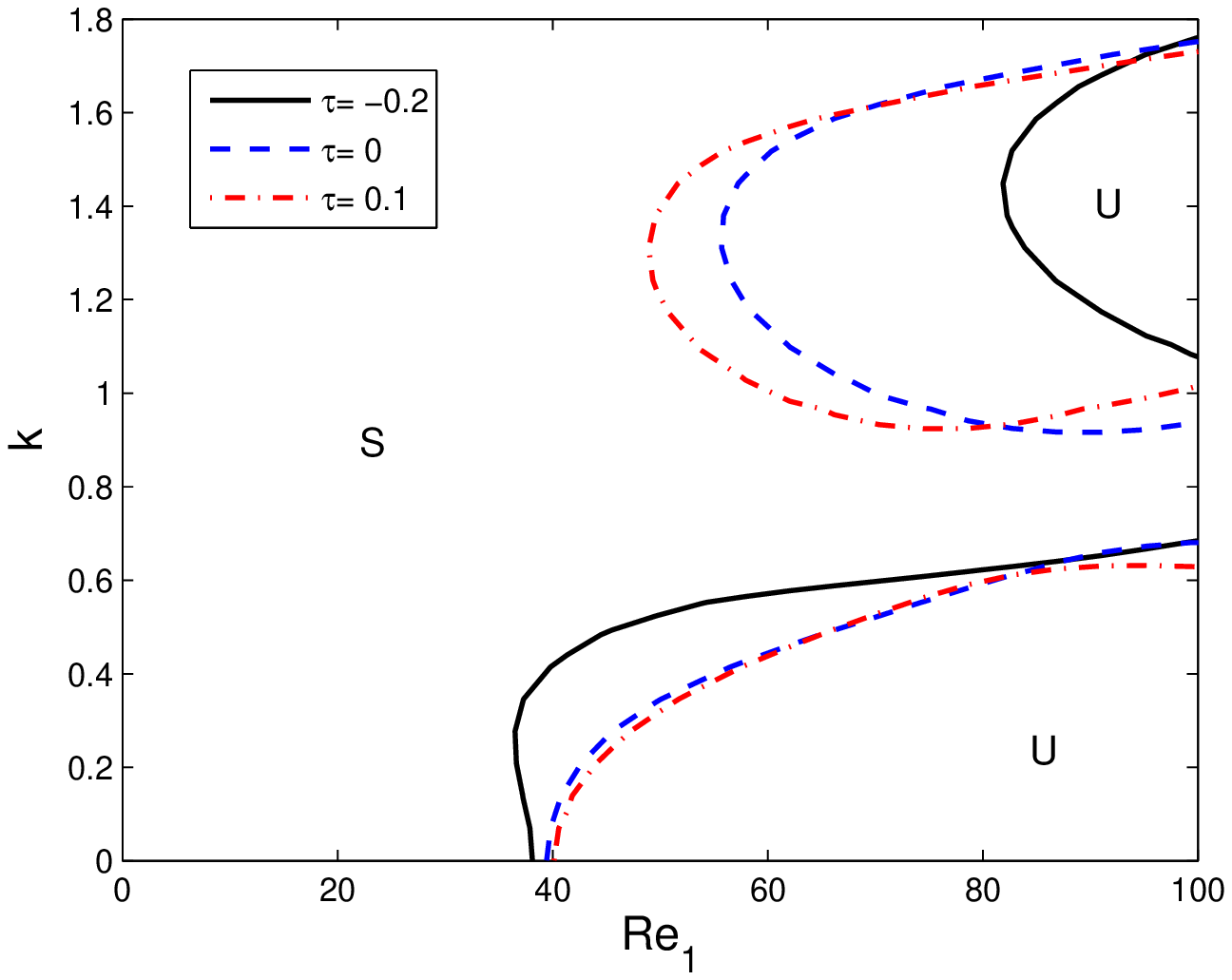}}
		\subfigure[$mr>1$]{\includegraphics*[width=7cm]{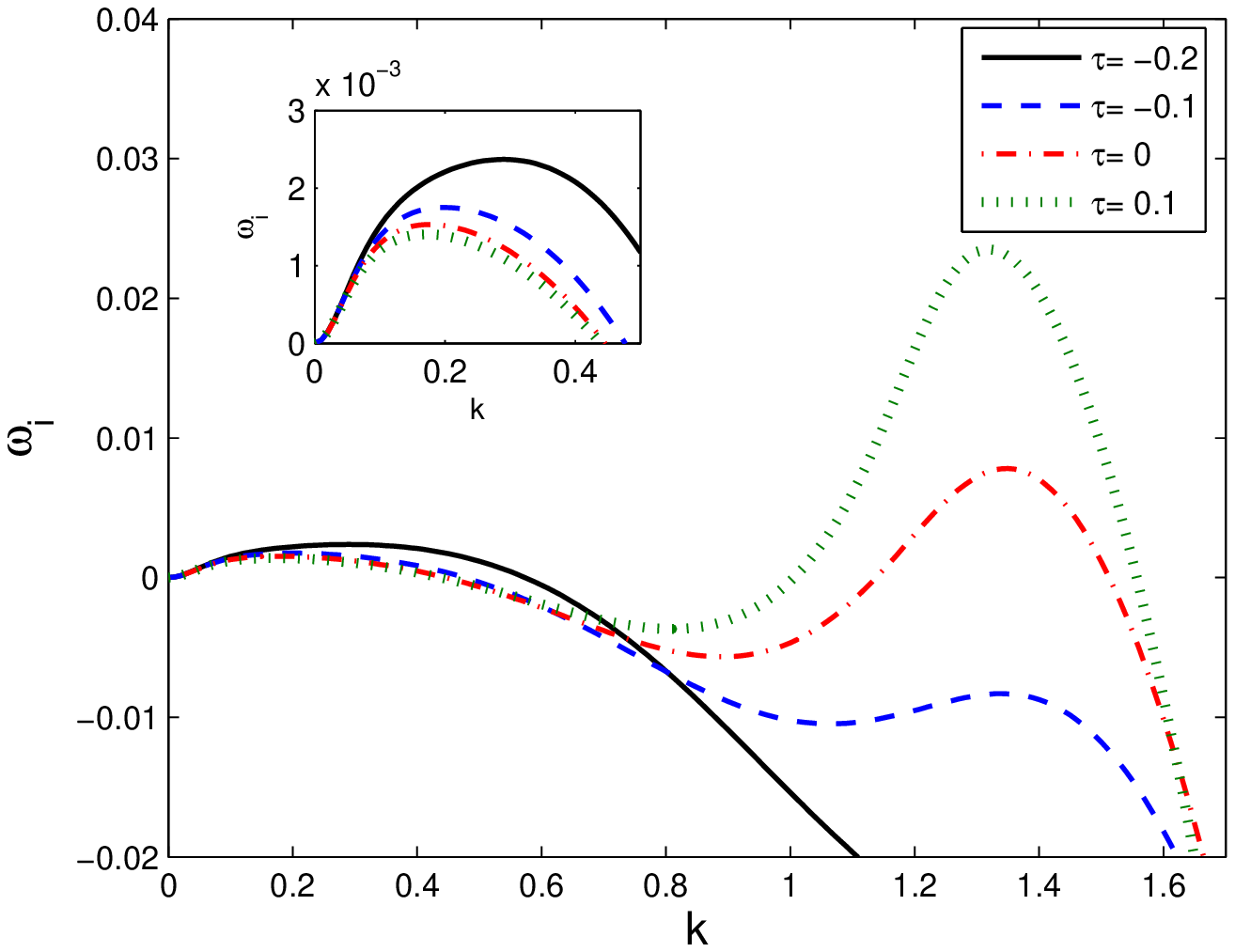}}
	\end{center}
	\caption{Marginal stability curve corresponding to varying external shear for (a) $mr>1$ with $r=1.1$ and (b) its corresponding scaled growth rate ($\omega_i$) as a function of wavenumber ($k$) with $Re_1=60$. The other constant flow parameters are $m=2.5$, $Pe_1=Pe_2=10000$, $Ca_1=Ca_2=1$, $\beta=0.04$, $\theta=0.2$ rad, $\delta=1$, $m=2.5$ and $Ma_1=Ma_2=0.1$.}\label{f7}
\end{figure}
\begin{figure}[h!]
	\begin{center}
		\subfigure[$mr<1$]{\includegraphics*[width=7cm]{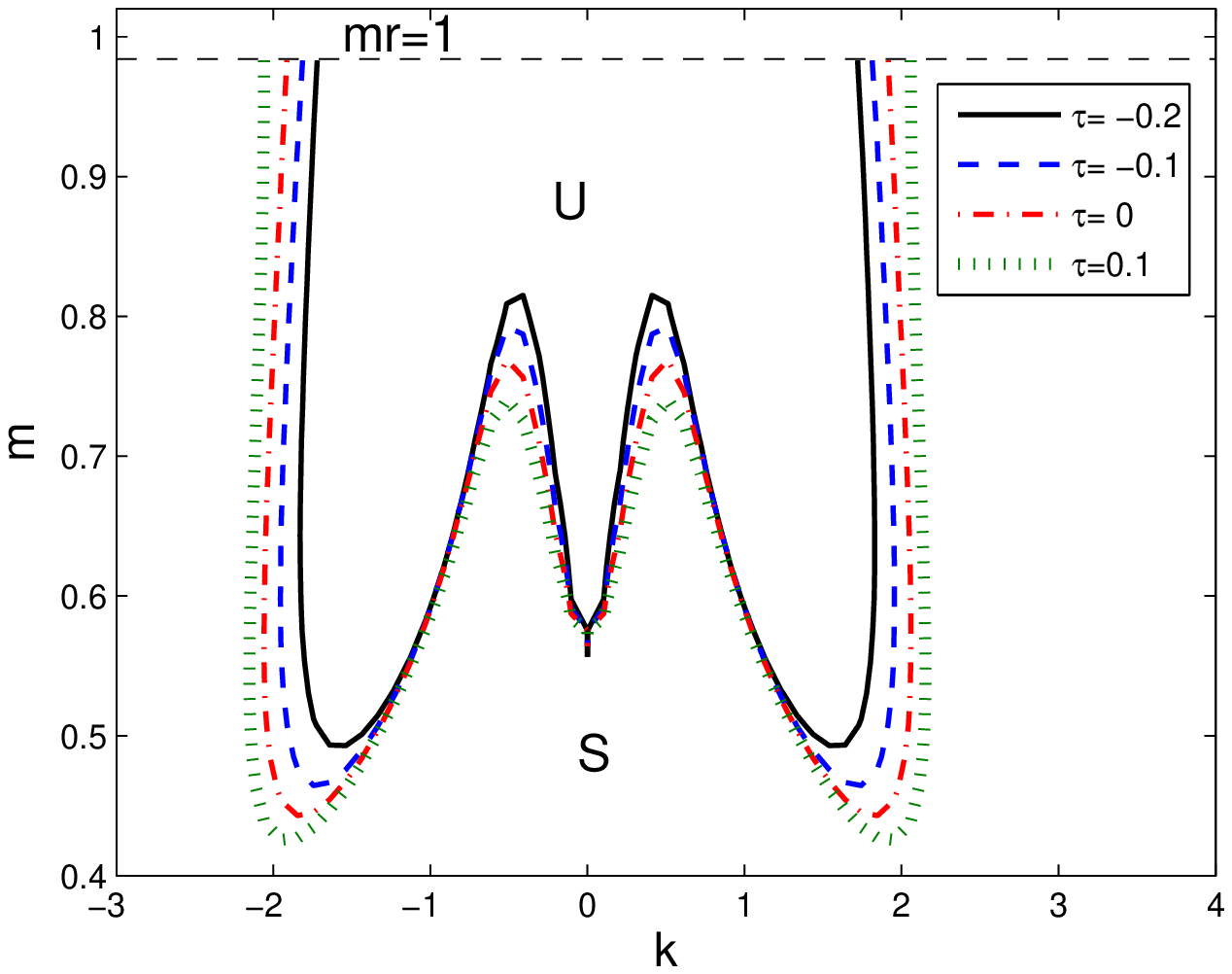}}
		\subfigure[$mr>1$]{\includegraphics*[width=7cm]{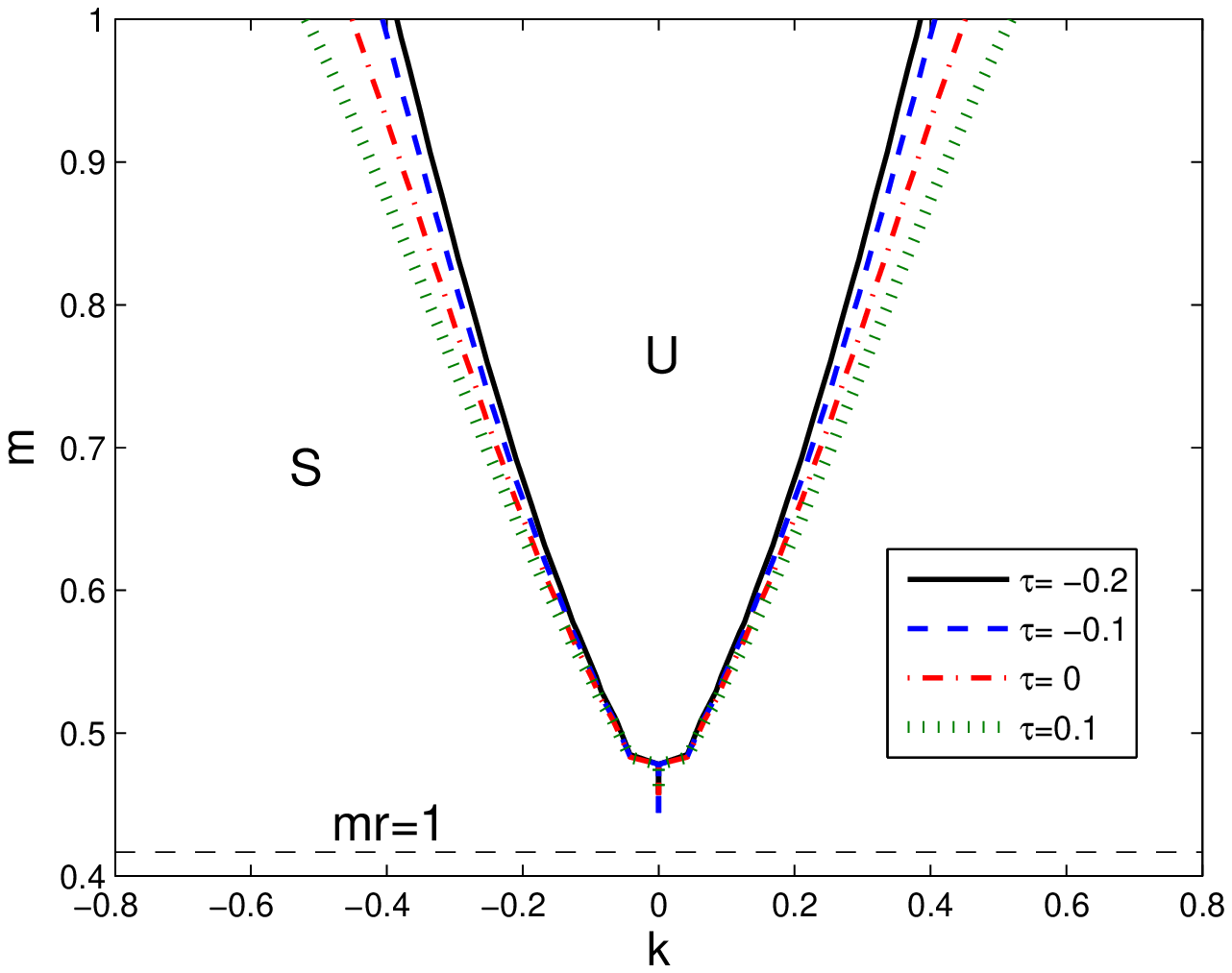}}
	\end{center}
	\caption{Marginal stability curve of the interface mode in $(m,k)$ plane for (a) $mr<1$ ($r=1.2$, $Re_1=25$) and (b) $mr>1$ ($r=2.4$, $Re_1=100$). The other flow parameters are $Pe_1=Pe_2=10000$, $Ca_1=Ca_2=1$, $\beta=0.04$, $\theta=0.2$ rad, $\delta=1$ and $Ma_1=Ma_2=0.1$.}\label{f8}
\end{figure}

The neutral stability curves show the effect of imposed shear for $mr>1$ provided that both $m>1$ and $r>1$ in Fig.~\ref{f7}(a). On comparing with $m<1$ as in Fig.~\ref{f4}(b), it is noticed that, in addition to the unstable interfacial mode bandwidth at the long-wave region, there exist a weaker unstable bandwidth at shorter-wave region. Also, the external shear plays dual role depending on the bandwidth of the unsatble waves. The critical $Re_1$ for the instability increases with higher value of $\tau$ in the long-wave region, whereas a reverse pattern is observed in the shorter-wave region. Further, the variation of unstable mode bandwidth is opposite in the case of $m>1$ (Fig.~\ref{f7}(a)) as compared to that of $m<1$ (Fig.~\ref{f4}(b)). In Fig.~\ref{f7}(b), the growth rate corresponding to the dominant unstable mode is plotted as a function of $k$ for different values of $\tau$ in the case of $m>1$. The growth rate decreases for increasing   value of $\tau$ in the long-wave region, however it follows reverse pattern for increasing value of external shear in the short-wave region. It is noticed that if the external shear applied opposite to the flow direction in the free surface surface, triggers more long-wave instability at the interface. On the other hand, the external shear increases the short-wave instability when it is applied along the direction of fluid flow.

\begin{figure}[h!]
	\begin{center}
		\subfigure[Long-wave region]{\includegraphics*[width=7cm]{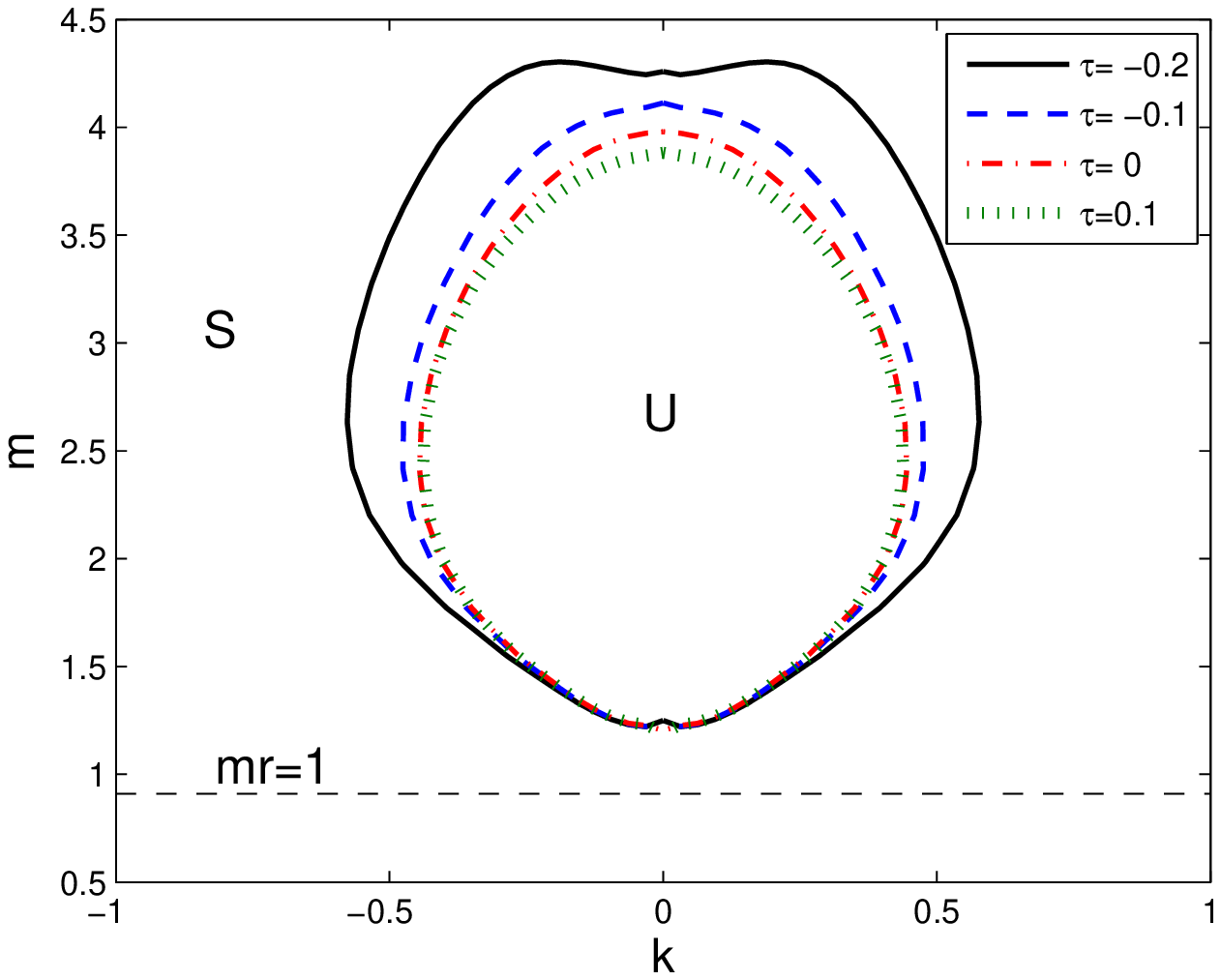}}
		\subfigure[Short-wave region]{\includegraphics*[width=7cm]{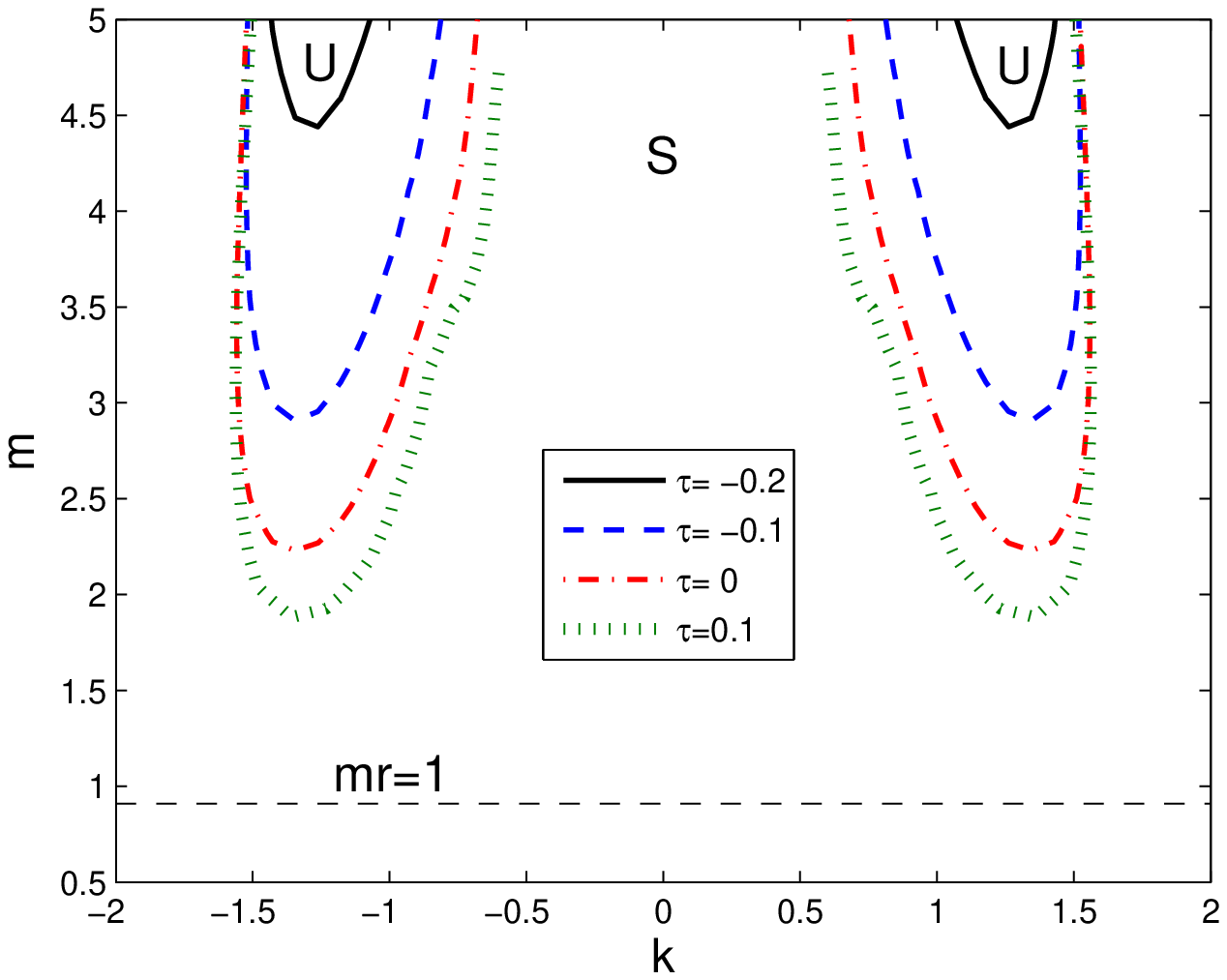}}
	\end{center}
	\caption{Marginal stability curve of the interface mode in $(m,k)$ plane for $mr>1$ 
		with $m>1$, $r=1.2$, $Re_1=60$, $Pe_1=Pe_2=10000$, $Ca_1=Ca_2=1$, $\beta=0.04$, $\theta=0.2$ rad, $\delta=1$ and $Ma_1=Ma_2=0.1$.}\label{f9}
\end{figure}
\begin{figure}[h!]
	\begin{center}
		\subfigure[]{\includegraphics*[width=7cm]{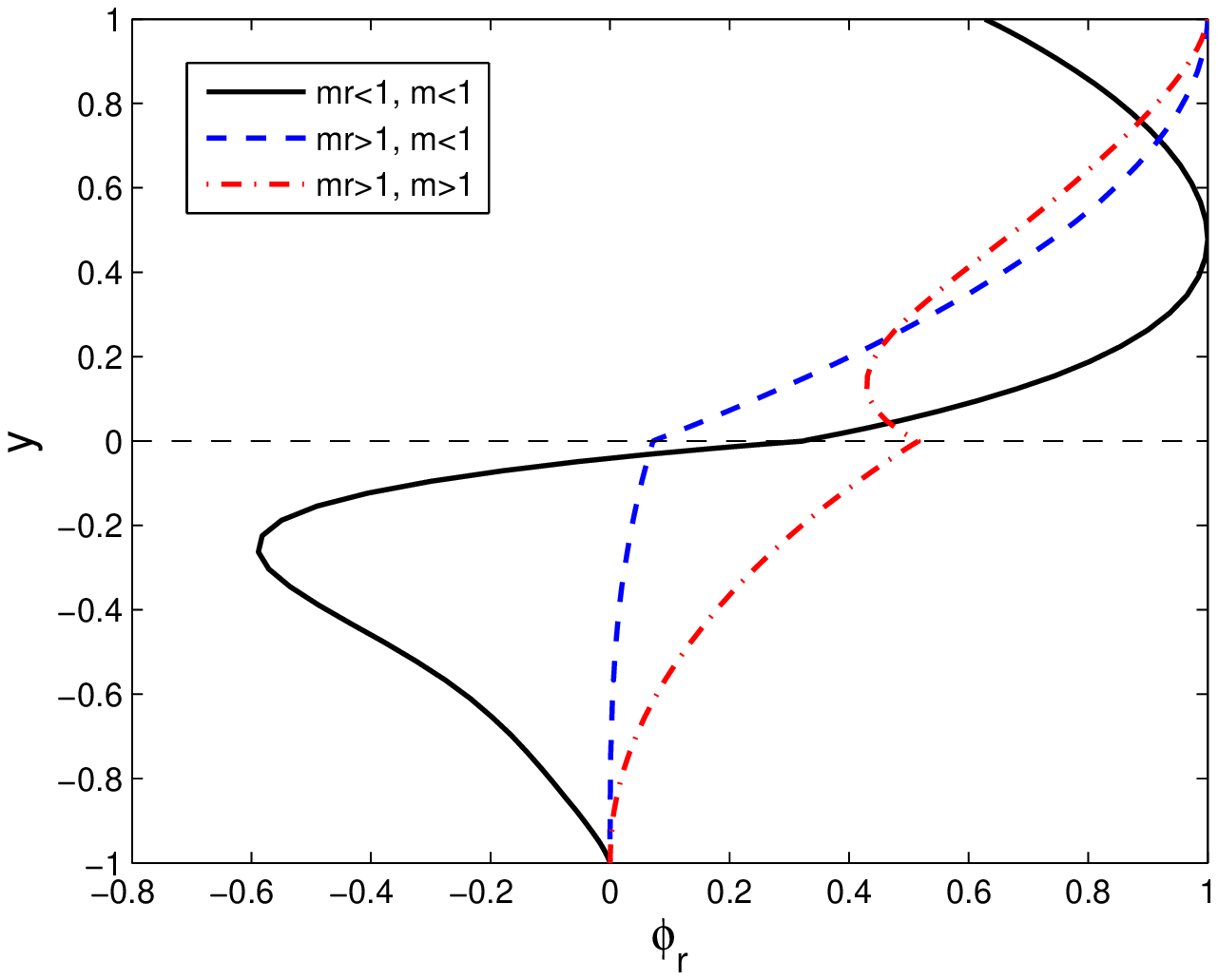}}
		\subfigure[]{\includegraphics*[width=7cm]{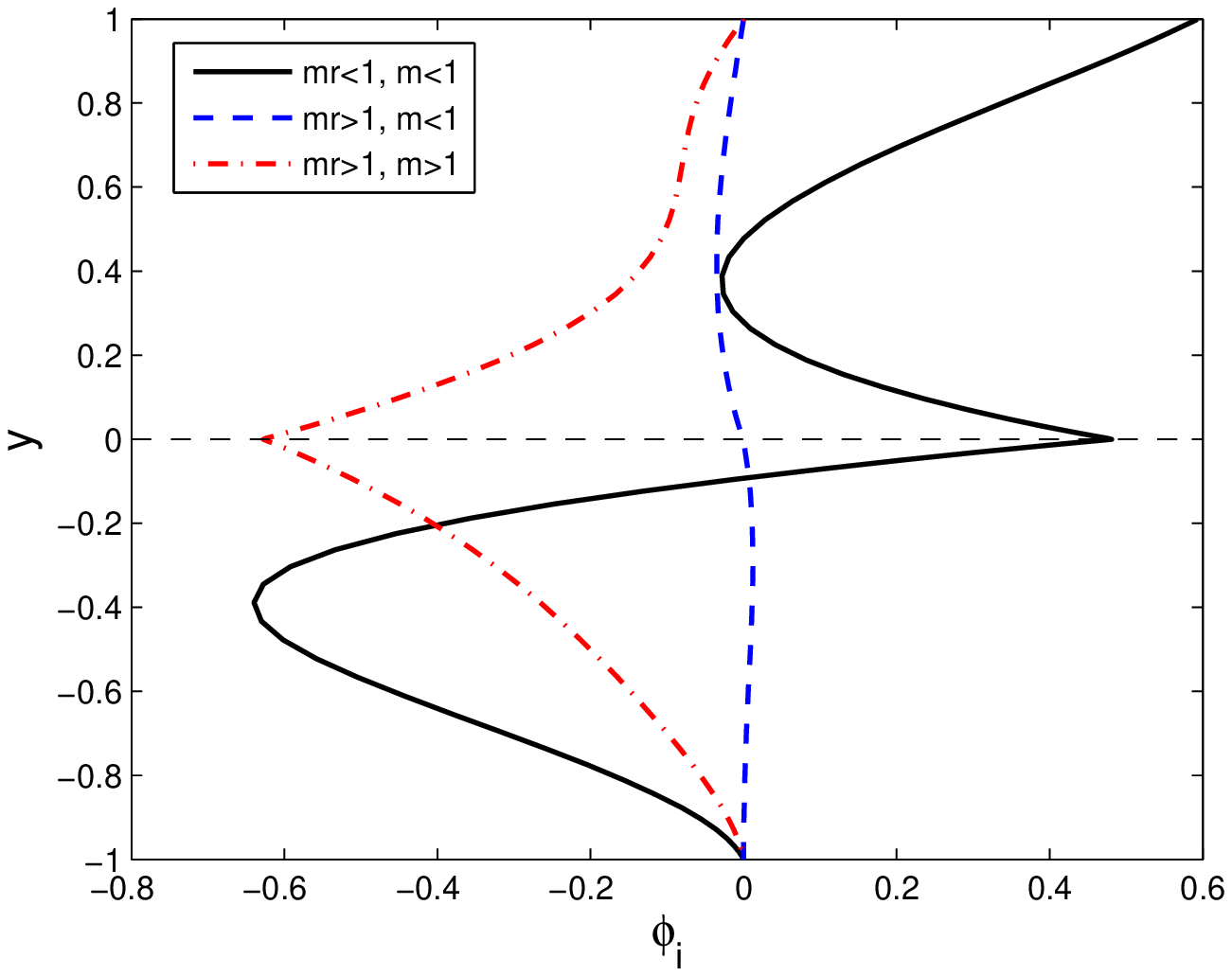}}
	\end{center}
	\caption{Eigenfunction ($\phi$) corresponding to the interface mode for various unstable zones with $Pe_1=Pe_2=10000$, $Ca_1=Ca_2=1$, $\beta=0.04$, $\theta=0.2$ rad, $\delta=1$ and $Ma_1=Ma_2=0.1$.}\label{f10}
\end{figure}

The marginal stability curve in the $(m,k)$ plane is plotted for the configurations, $mr<1$ and $mr>1$ in Figs.~\ref{f8}(a) and (b), respectively when a more viscous fluid is flowing on a less viscous ($m<1$). On the dashed line in both the figures, $m$ and $r$ satisfies $mr=1$ for the given value of the density ratio $r$. This line forms the critical bound for the interfacial instability, namely the upper and lower bounds for the marginal stability region in Figs.~\ref{f8}(a) and (b), respectively. For $mr<1$, the region of unstable mode bandwidth is more dominant near the long-wave region and less significant near the short-wave region. It is noticed that the present result validates the previous observation for $m=0.5$ as in Fig.~\ref{f8}(a), where there exist only short-wave instability. With increase in the values of $\tau$, the unstable mode bandwidth increases. In the case of $mr>1$ from Fig.~\ref{f8}(b), the unstable mode bandwidth increases for increase in the values of $m$. Further, it is observed that the unstable mode bandwidth increases with stronger  external shear $\tau$. In both the figures, it is perceived that the critical condition of interfacial instability attained only one time for $k=0$.

In Figs.~\ref{f9}(a) and (b), the bandwidth of unstable modes in both long- and short-wave regions are plotted respectively for $mr>1$ provided that $m>1$. In the case of long-wave region, the neutral conditions are attained two-times for a given Reynolds number. This depicts that there exist a sub-critical instability at long wave region while considering the high viscous fluid in the lower layer. With increase in the value of $\tau$, the unstable mode bandwidth decreases which is validated with the previous observation in Fig.~\ref{f7}(b). For the short-wave region as in Figs.~\ref{f9}(b), the unstable mode bandwidth increases for increase in the external shear. Further, there is no occurrence of neutral condition in the case of short-wave region. The observation are similar to the scales growth rate result for a fixed value of $m$ greater than unity (Fig.~\ref{f7}(b)).

In Figs.~\ref{f10}(a) and (b), the real and imaginary part of eigenfunction, respectively, corresponding to the interface mode are plotted for three different unstable zones namely, (i) zone 1 ($mr<1$, $m<1$), (ii) zone 2 ($mr>1$, $m<1$) and (iii) zone 3 ($mr>1$, $m>1$). The interfacial mode shape undergoes different variation in the three unstable zones under the influence of external shear. At the interface from Fig.~\ref{f10}(a), the velocity of fluid under zone 3 is more as compared to other two zones resulting in the higher value of $\phi_r$ for the zone 3. However, in the case of $\phi_i$ as in Fig.~\ref{f10}(b), the $\phi_i$ corresponding to the zone 3 has the lowest value as compared to other zones at interface.

\subsection{Effect of imposed shear on the surface mode}

\begin{figure}[h!]
	\begin{center}
		\subfigure[]{\includegraphics*[width=7cm]{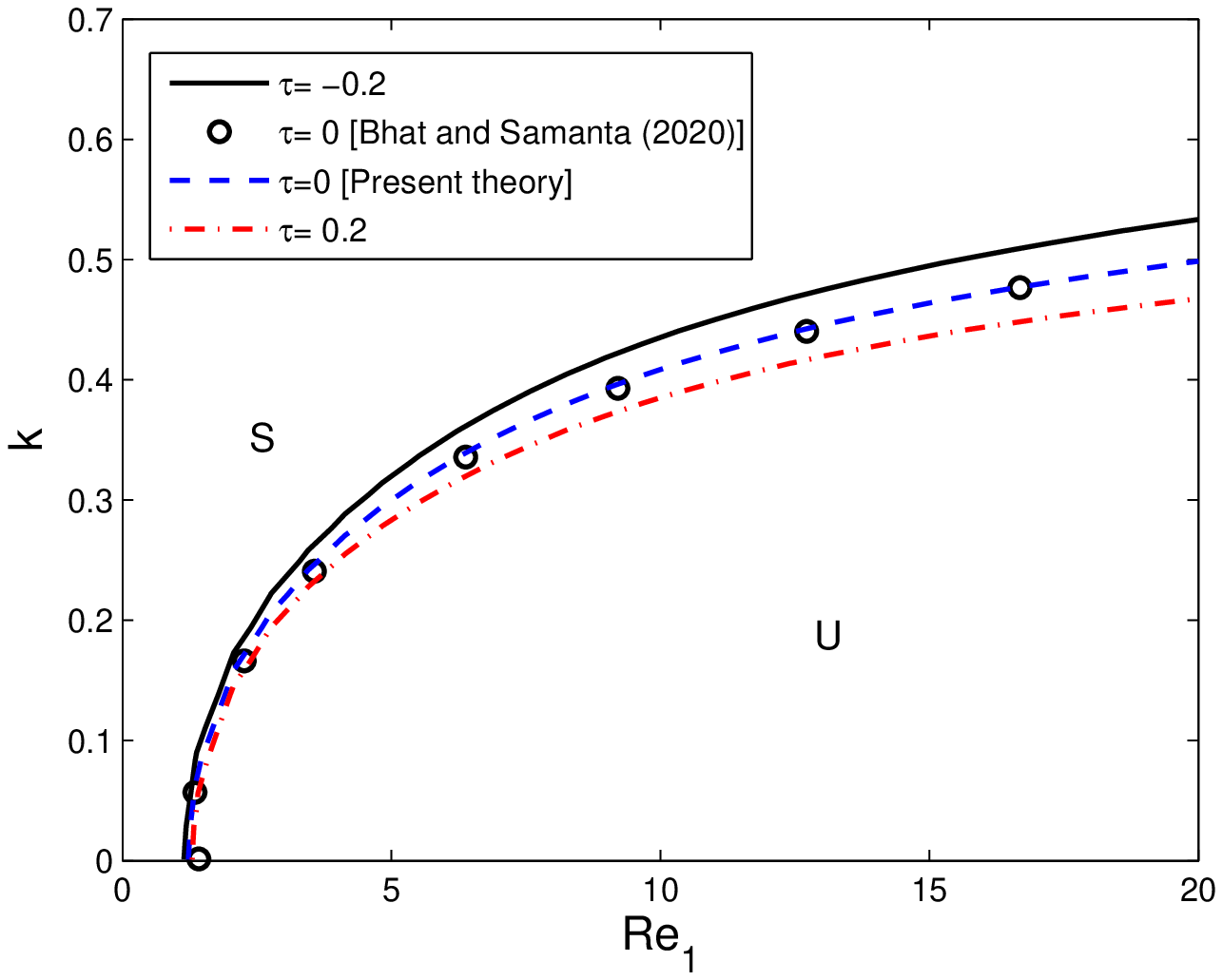}}
		\subfigure[]{\includegraphics*[width=7cm]{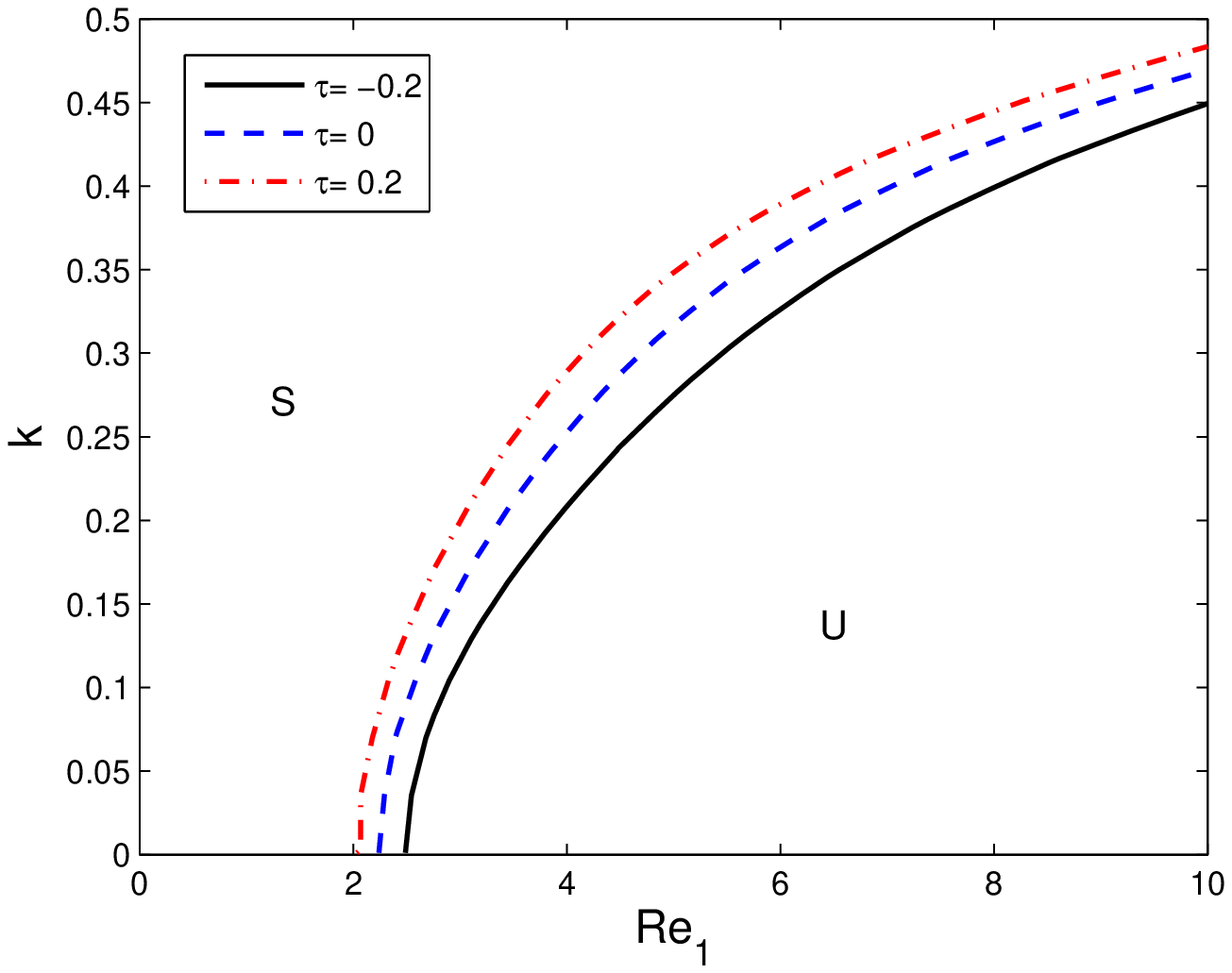}}
	\end{center}
	\caption{(a) Neutral stability curve corresponding to varying external shear ($\tau$) when the fixed flow parameters are $Pe_1=Pe_2=10000$, $Ca_1=Ca_2=1$, $\beta=0.04$, $\theta=0.2$ rad, $\delta=r=1$, $m=0.5$, $r=1.1$ and $Ma_1=Ma_2=0.1$, and (b) Stability boundaries for limiting single fluid falling film with imposed shear (\cite{bhat2019linear}). 
	}\label{f11}
\end{figure}

In this subsection, the influence of external shear on the surface mode instability is investigated for different set of flow parameters. In Fig.~\ref{f11}(a), the marginal stability curves are plotted in $(Re_1,k)$ plane for various imposed shear rate. It is noticed that the present theory is in well-agreement with the existing result of \cite{bhat2020linear} in the case of two-layer film flow down the slippery plane in the absence of imposed external shear at the free surface (for $\tau=0$). Further, our results confirms that the external shear applied in the direction of film flow reduces the bandwidth of the  unstable surface mode bandwidth and on applying the external shear in the opposite direction to the film flow strengthens the bandwidth of surface mode instability. 
Thus, in the case of considered two-layer flow, the influence of imposed shear on the surface mode is exactly opposite to that for the shear imposed surfactant laden single layer falling film (\cite{bhat2019linear}), where the surface unstable mode bandwidth increases/decreases for positive/negative values of external shear rate (see Fig.~\ref{f11}(b). It is really interesting to see that presence of two different fluid layer with density/viscosity variation completely changing the stability mechanism of imposed shear. Note that, on applying the external shear in the flow direction to the stratified flow system, the velocity of the upper layer fluid increases near the surface and consequently, there is a decrease in the velocity of lower layer as well as interfacial velocity. However, the change in interfacial shear overcomes the shear rate at free surface region, which cases an hindrance to the inertia force on the convective flow and yields a stabilizing effect on the surface mode. By contrast, if the external shear stress acts opposite to the streamwise flow direction, a strong back-flow phenomena occurs at the free surface region, which enhances the surface mode instability. 
\begin{figure}[h!]
	\begin{center}
		\subfigure[]{\includegraphics*[width=7cm]{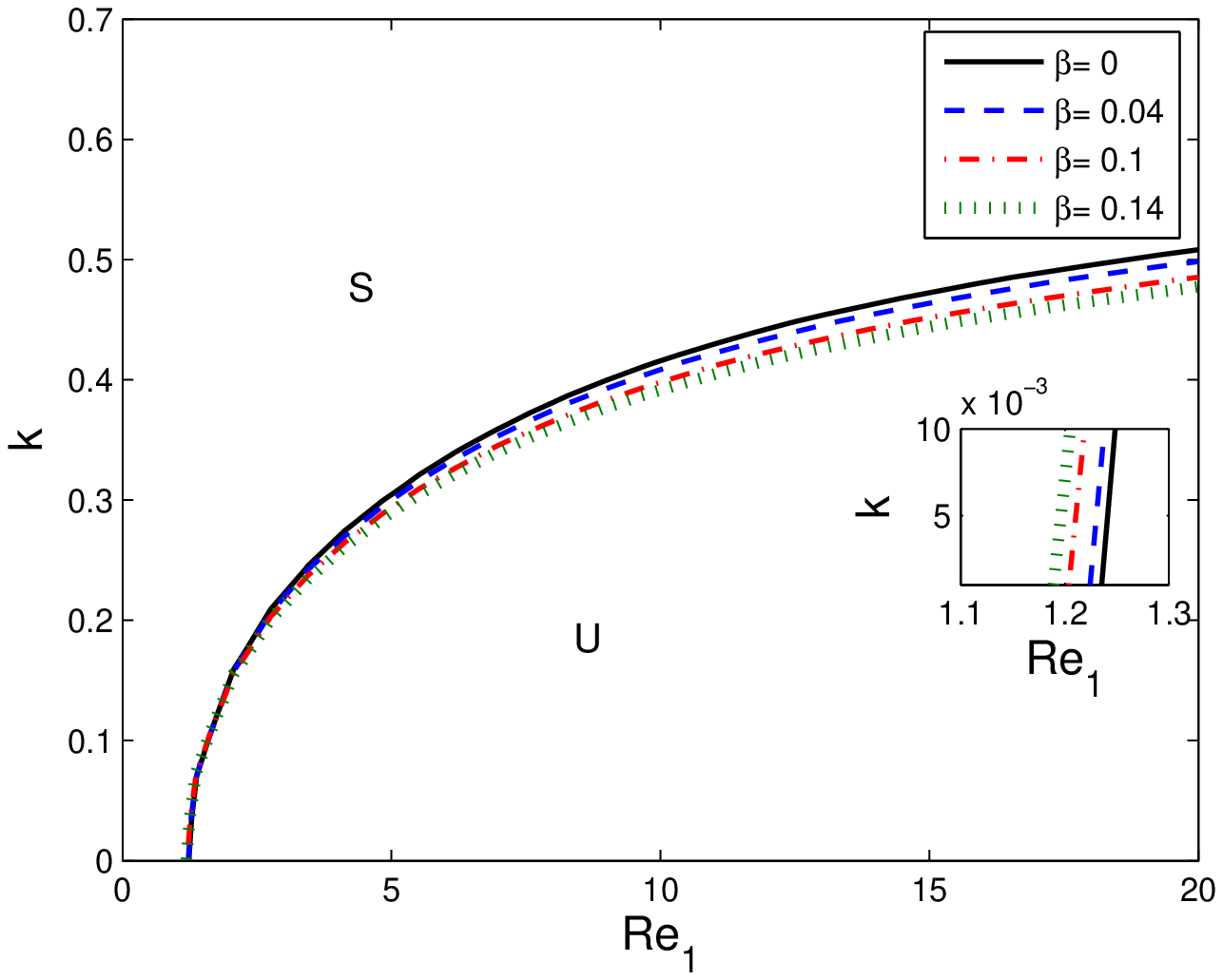}}
		\subfigure[]{\includegraphics*[width=7cm]{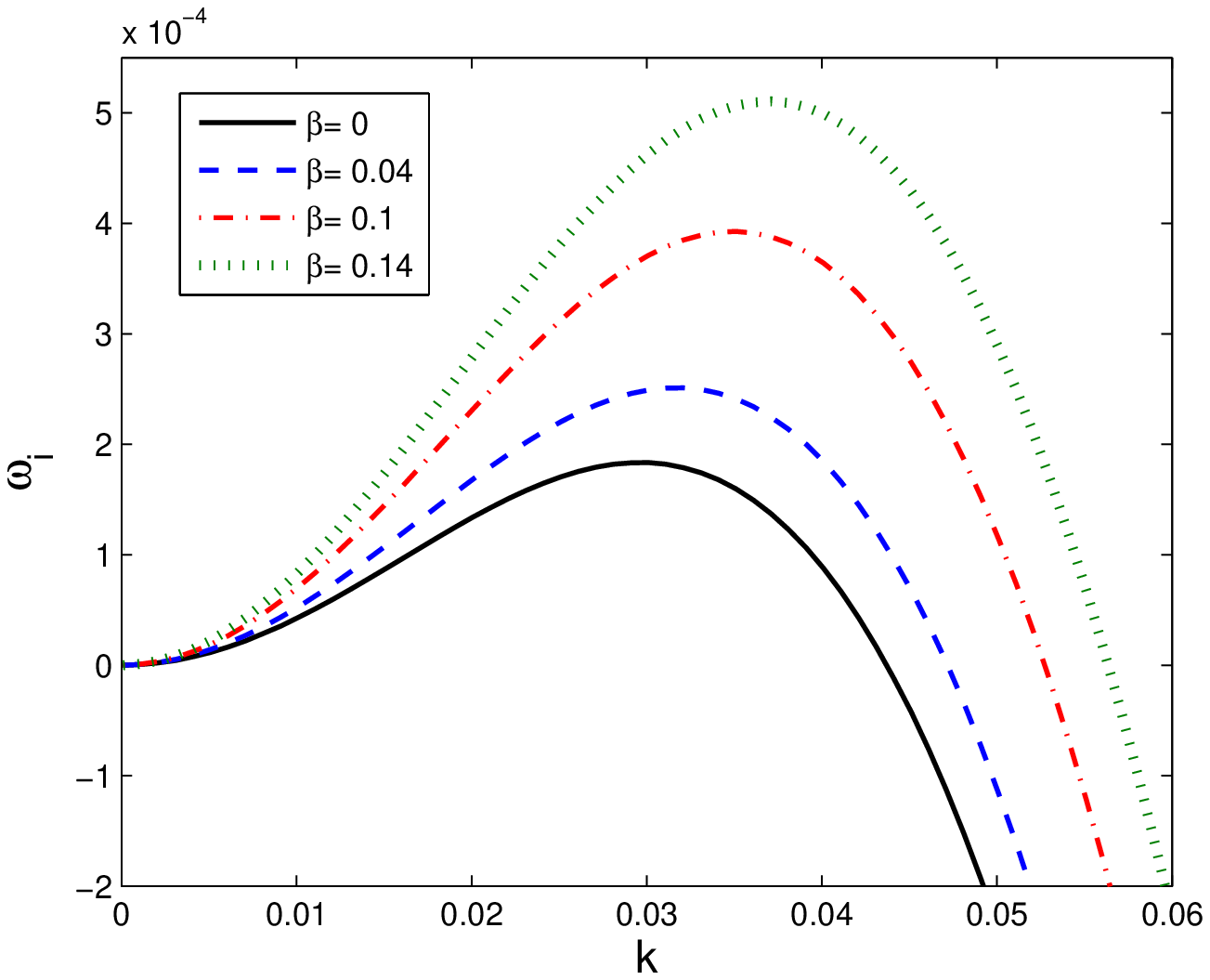}}
	\end{center}
	\caption{(a) Marginal stability curve corresponding to varying slip length ($\beta$) and (b) its corresponding scaled growth rate ($\omega_i$) as a function of wavenumber ($k$) with $Re_1=1.3$. The other constant flow parameters are $Pe_1=Pe_2=10000$, $Ca_1=Ca_2=1$, $\tau=0$, $\theta=0.2$ rad, $\delta=r=1$, $m=0.5$, $r=1.1$ and $Ma_1=Ma_2=0.1$.}\label{f12}
\end{figure}

In Fig.~\ref{f12}(a), the effect of slip length on the marginal stability curve is illustrated in the $(Re_1,k)$ plane in the absence of external shear. The neutral curve behaves differently for very smaller values of $k$ as shown in inset plot of Fig.~\ref{f12}(a) as compared to the larger values of $k$. A dual role of wall velocity is observed similar to the case of single fluid falling film. For the onset of instability $k\in [0, 0.06]$, there is a mild destabilization by the reduction in the critical Reynolds number with the increasing value of slip parameter, whereas the surface mode bandwidth decreases with larger $\beta$ for shorter-waves.
The wall shear rate changes with the variation of the slip length, which in turn affects the interfacial and surface velocity of the base flow. Further, when inertial effect is strong enough the surface instability stabilizes gradually for increase in the value of $\beta$.
In order to validate the destabilizing effect of slip parameter on the long waves as seen in Fig.~\ref{f12}(a), the corresponding growth rate is plotted for the region  $0<k<0.06$ in Fig.~\ref{f12}(b) and it is observed that the growth rate of surface instability accelerates for larger values of slip length.

\begin{figure}[h!]
	\begin{center}
		\subfigure[]{\includegraphics*[width=7cm]{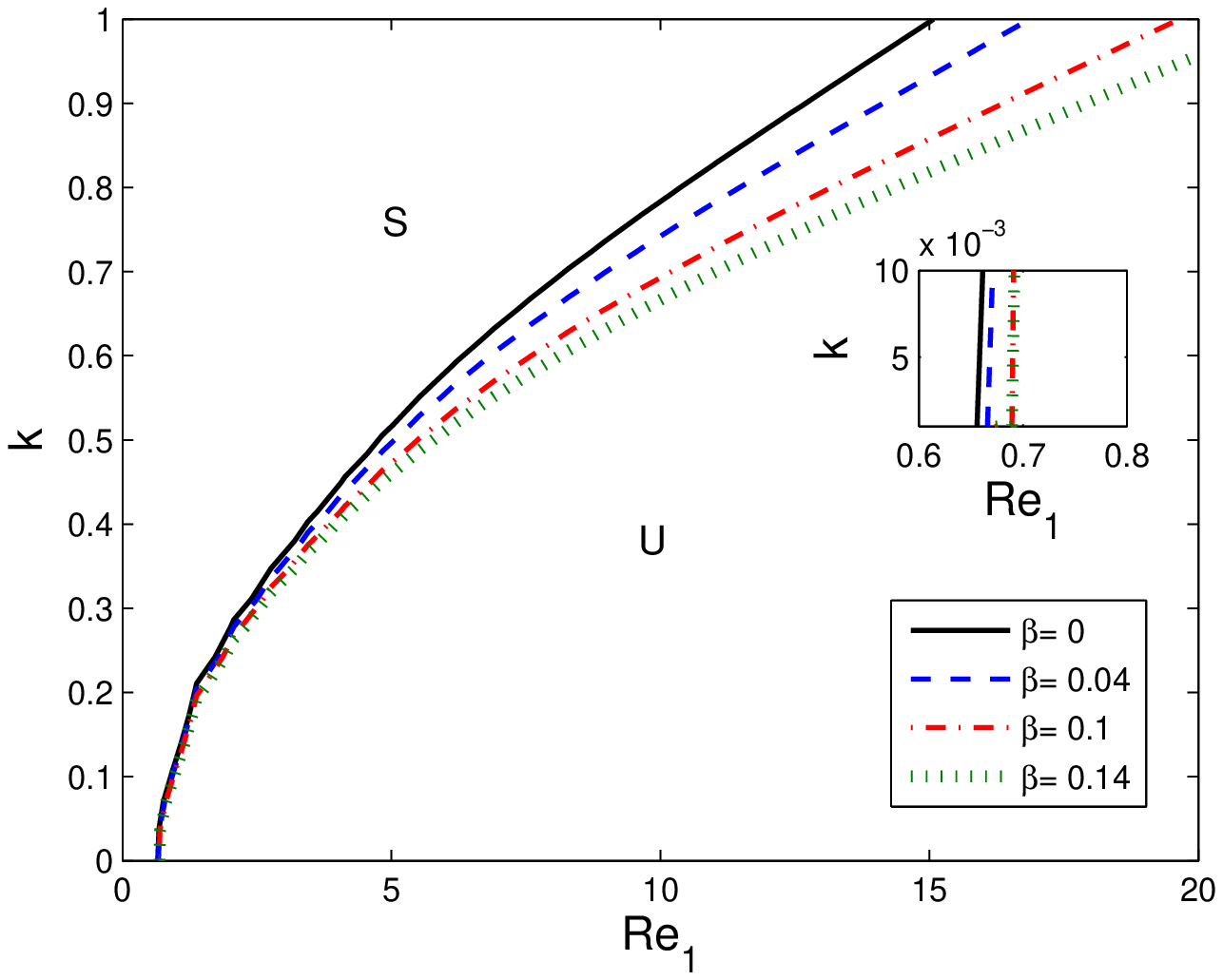}}
		\subfigure[]{\includegraphics*[width=7cm]{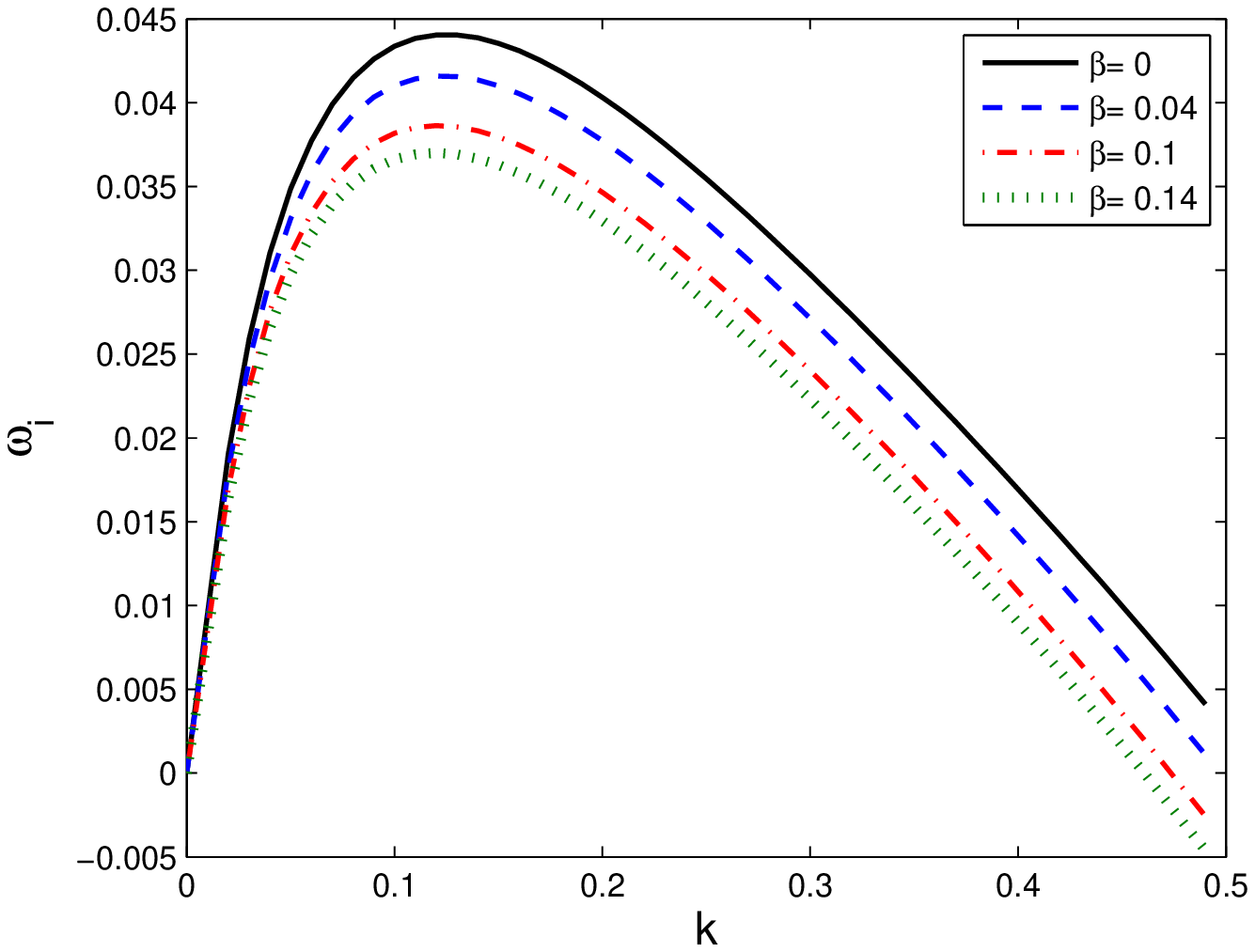}}
	\end{center}
	\caption{(a) Neutral stability curves corresponding to varying slip length ($\beta$) and (b) its corresponding scaled growth rate ($\omega_i$) as a function of wavenumber ($k$) with $Re_1=5$. The other constant flow parameters are $Pe_1=Pe_2=10000$, $Ca_1=Ca_2=1$, $\tau=-1$, $\theta=0.2$ rad, $\delta=r=1$, $m=0.5$, $r=1.1$ and $Ma_1=Ma_2=0.1$.}\label{f13}
\end{figure}

The role of wall slip in the presence of imposed shear is captured in Fig.~\ref{f13}. The neutral stability curves in the $(Re_1,k)$ plane and its corresponding growth rate as a function of $k$ are analyzed for the same flow parameters as considered in Fig.~\ref{f12} after including a strong external shear effect ($\tau = -1$). Here, the external shear is applied opposite to the direction of two-layer film flow (i.e. for negative values of $\tau$). It is observed from Fig.~\ref{f13}(a) that the unstable mode bandwidth decreases for increase in the value of $\beta$ for all values of $Re_1$. There is no longer any destabilizing effect of the wall velocity slip close to the criticality (in particular for $k\to 0$). It fact ensures that slip parameter supplies only stabilizing effect on the surface mode instability, when external shear is included. This is possibly because of the energy balancing between the extra imposed shear at the free surface to the interfacial tension. The corresponding growth rate as a function of wavenumber for different values of $\beta$ is plotted in Fig.~\ref{f13}(b). It is noticed that the temporal growth rate of the surface instability decreases for stronger velocity slip which advances the wall shear rate.
\begin{figure}[h!]
	\begin{center}
	\subfigure[]{\includegraphics*[width=7cm]{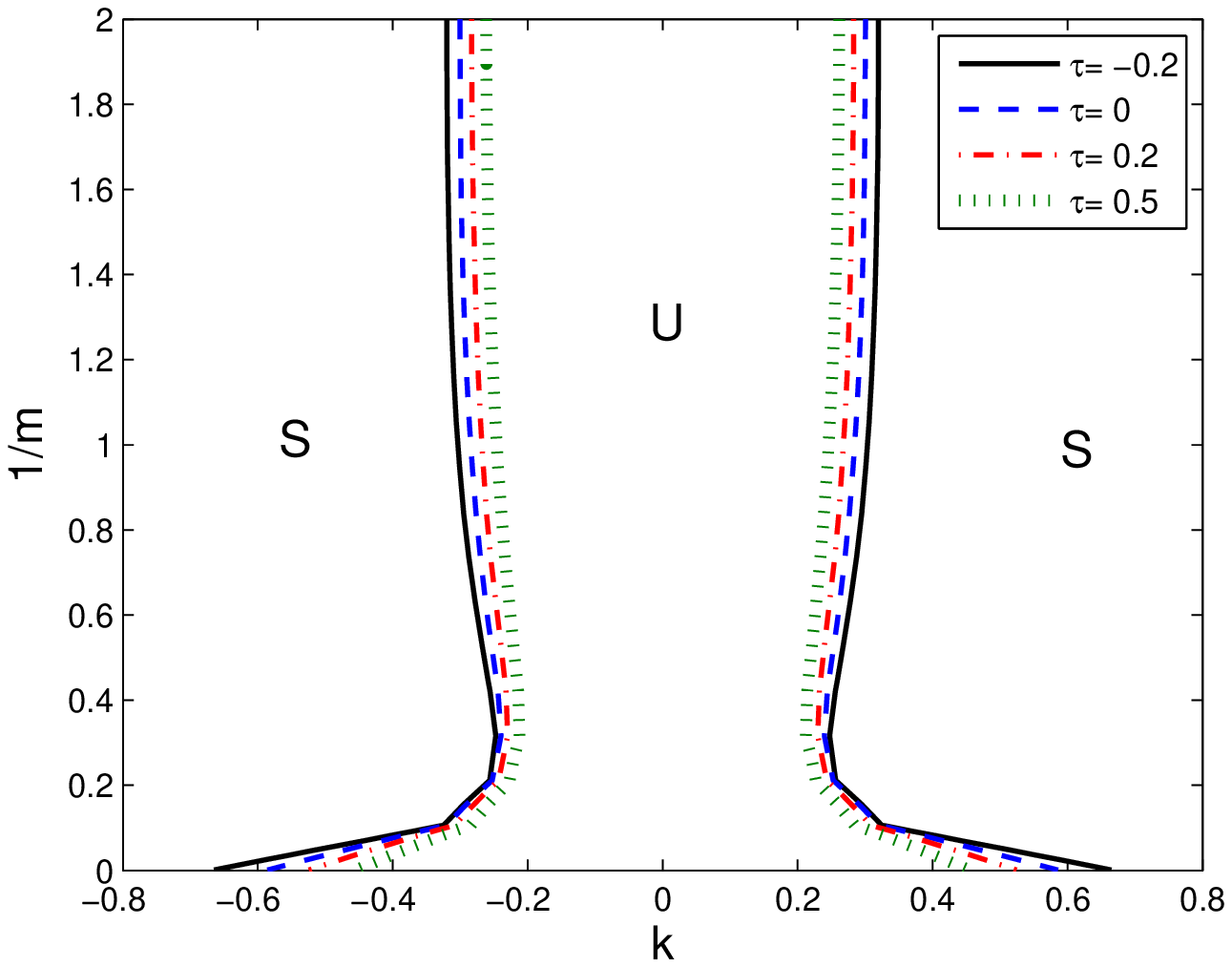}}
	\subfigure[]{\includegraphics*[width=7cm]{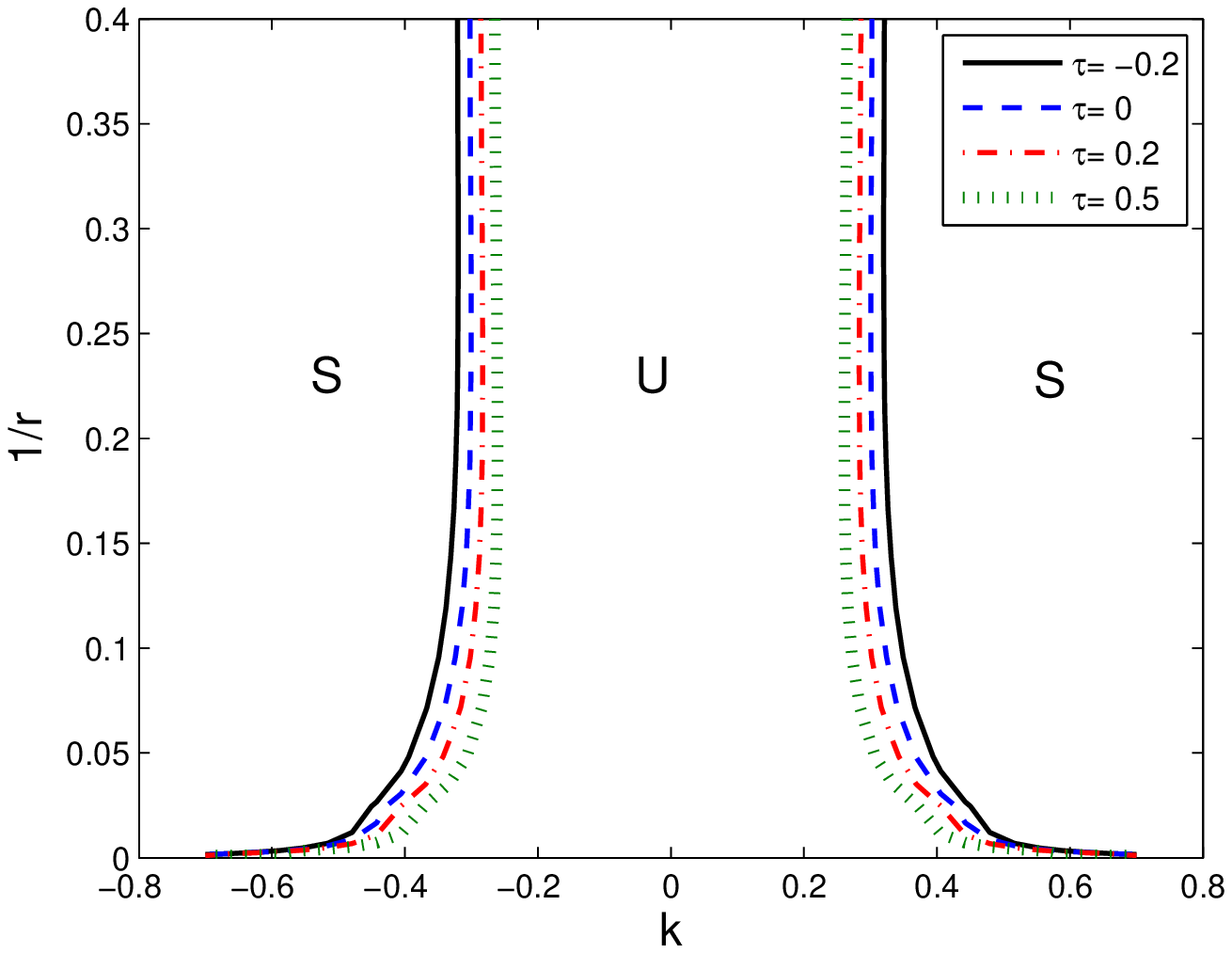}}
	\end{center}
	\caption{Marginal stability curves: (a) in the $(1/m,k)$ plane with $r=1.1$ and (b) in the $(1/r,k)$ plane with $m=0.5$ for different values of $\tau$ when $Re_1=5$, $Pe_1=Pe_2=10000$, $Ca_1=Ca_2=1$, $\theta=0.2$ rad, $\delta=1$ and $Ma_1=Ma_2=0.1$.}\label{f14}
\end{figure}

\begin{figure}[h!]
	\begin{center}
		\subfigure[]{\includegraphics*[width=7cm]{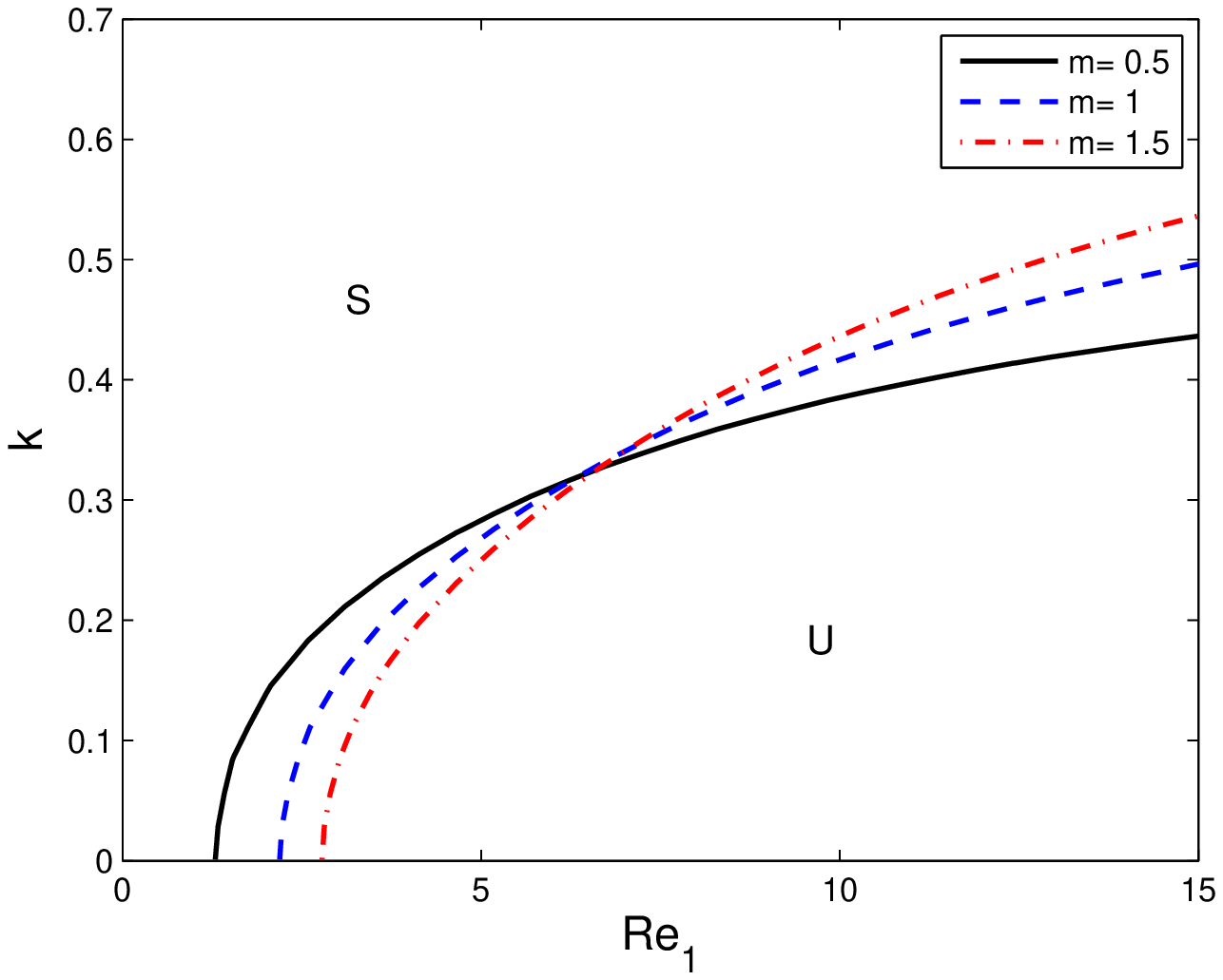}}
		\subfigure[]{\includegraphics*[width=7cm]{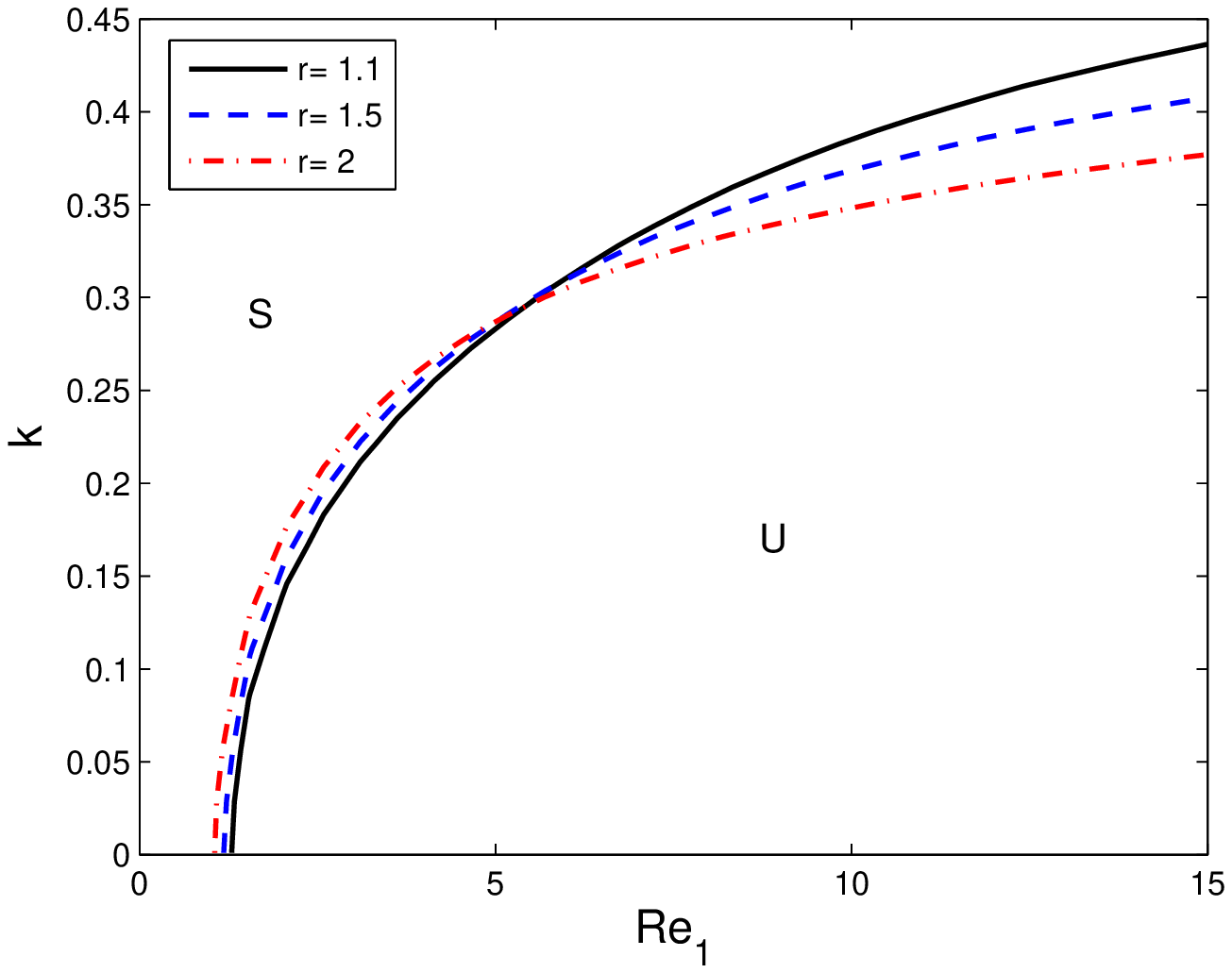}}
	\end{center}
	\caption{Effects of viscosity and density stratification on the neutral stability curves on the surface mode: (a) for varing viscosity ratio $(m)$ with $r = 1.1$ and (b) for varing density ratio $(r)$ with $m = 0.5$. The other flow parameters are $Pe_1=Pe_2=10000$, $Ma_1=Ma_2=0.1$,  $Ca_1=Ca_2=1$, $\delta=1$, $\tau=0.2$ and $\beta=0.04$.}\label{f15}
\end{figure}

The Marginal stability curves are drown in Fig.~\ref{f14}(a) and (b), respectively, in the $(1/m,k)$ and $(1/r,k)$ plane for exploring the substantial effect of external shear on surface mode. It is observed that, the unstable mode bandwidth is prevalent near the small wavenumber range, which gradually increases for lower values of $1/m$. When the viscosity in the upper layer fluid is much smaller as compared to the lower layer fluid, the long-wave instabilities are more significant. Further, with increase in the viscosity of upper layer, the long-wave instability corresponding to the surface mode diminishes initially and remains constant for larger values of $\mu_1$. On increasing the external shear, the unstable mode bandwidth decreases owing to the balancing of the shear rate at the free surface and the shear layer near the interface. This happens as a consequence of momentum conservation in the two-layer fluid system. From Fig.~\ref{f15}(a), it is found that the bandwidth of unstable modes is more significant near the small wavenumber region and it gradually decreases, then remains invariant for increasing the value of $1/r$.  The effect of density ratio on the marginal stability curve is more significant for the values of $1/r \in [0,0.01]$, which implies that the net driving force in the upper layer fluid increases due to the presence highly dense lower layer fluid, constituting more surface instabilities. With increase in the values of $1/r$, the density of upper layer fluid increases, which decreases the surface instabilities to some extent. However, there is no effect of upper layer density on the surface instability of two layer fluid for larger values of $1/r$. This occurs due to the balanced net driving force between the upper and lower fluids as the value of $r$ is approaching towards the unity along with the negligible effect of inertia on the unstable surface modes.

In Figs.~\ref{f15}(a) and (b), the neutral stability boundaries are plotted to describe the influence of viscosity and density ratio on the surface mode. The numerical results unveil that both the viscosity and density ratio play a dual role in the surface instability. It is observed that the unstable region corresponding to the surface mode deceases close to the threshold of instability, while it increases away from the threshold of instability after a critical value of wave number as soon as the viscosity ratio magnifies. In fact, the viscous shear stress of the upper layer enhances with the higher value of viscosity ratio to balance the same of the lower layer at the interface that causes a slower flow of the upper fluid layer and yields a stabilizing effect when inertial effects are not strong enough. Whereas, at the moderate wave numbers when the inertial effects dominants over the viscous effects a destabilizing impact of viscosity stratification is seen. However, the effect of density ratio on the surface mode instability is completely opposite to that of the viscosity ratio (Fig.~\ref{f15}(b)). In particular, for $r>1$ the lower layer fluid becomes more dense than the upper layer and it makes slower the impact of
the driving force of the heavier lower fluid layer because of the larger magnitude of the depthwise gravitational force. 

\begin{figure}[h!]
	\begin{center}
		\subfigure[]{\includegraphics*[width=7cm]{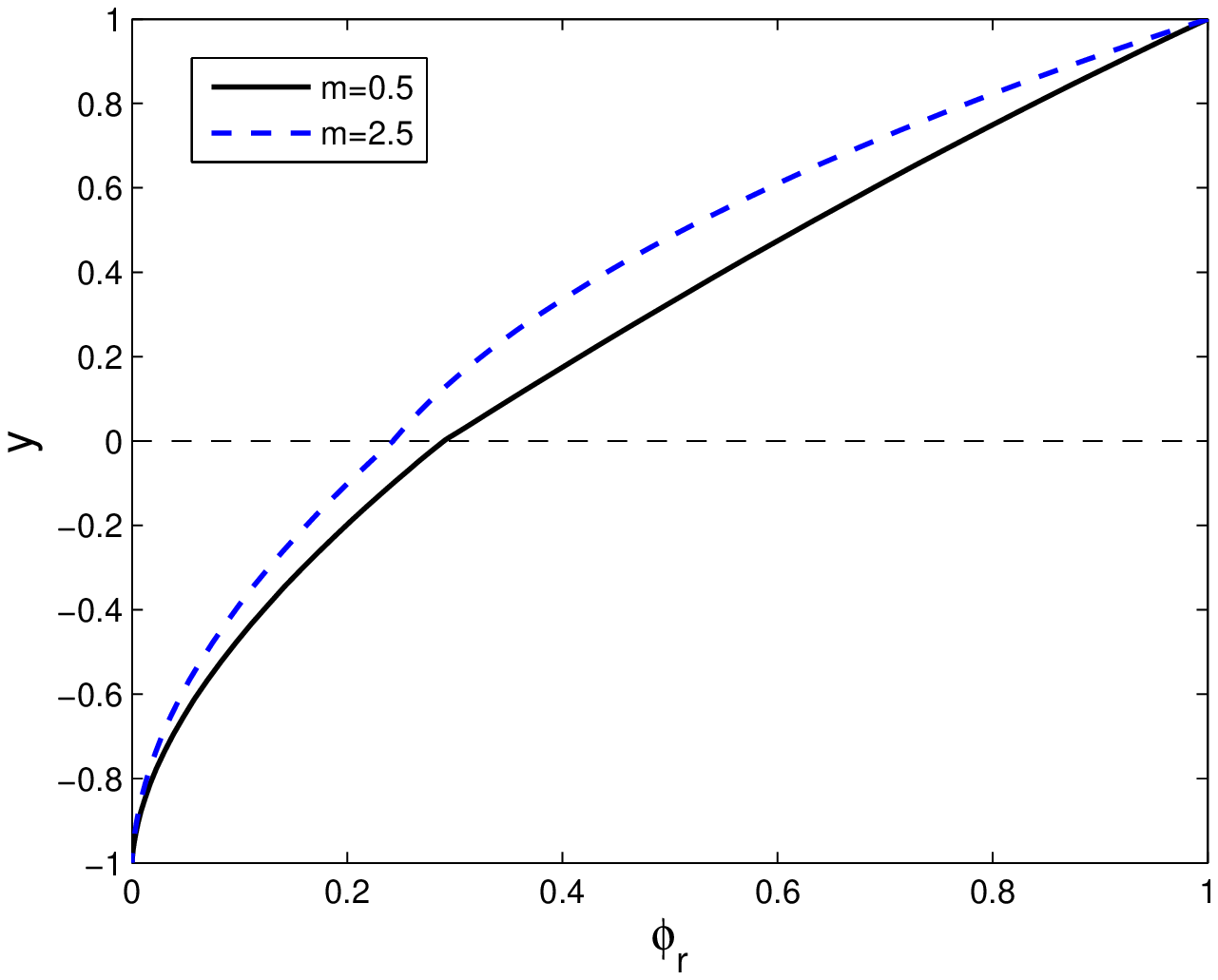}}
		\subfigure[]{\includegraphics*[width=7cm]{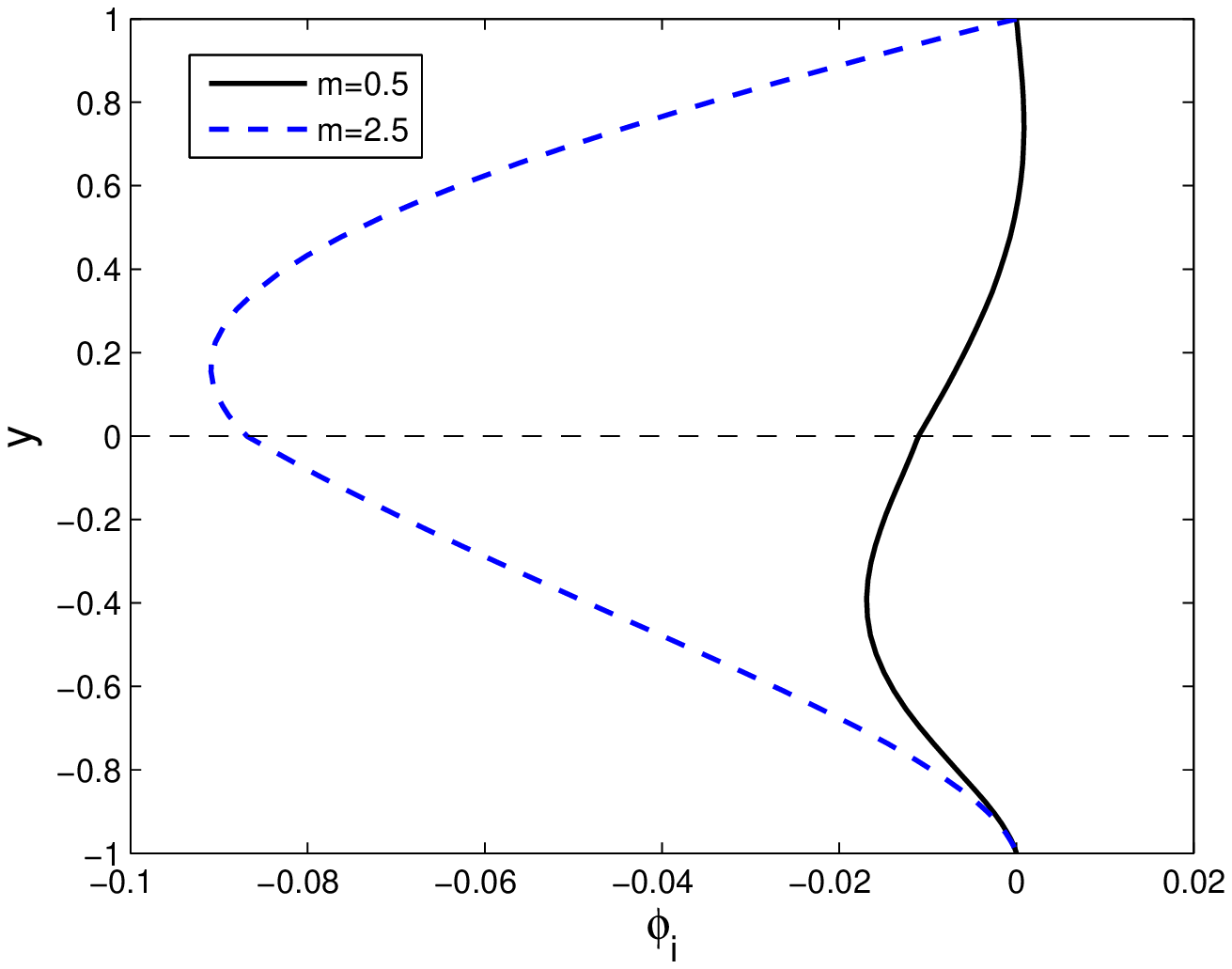}}
	\end{center}
	\caption{Eigenfunction ($\phi$) corresponding to the surface mode for various $m$ with $r=1.1$, $Re_1=5$, $\tau=0.1$, $Pe_1=Pe_2=10000$, $Ca_1=Ca_2=1$, $\beta=0.04$, $\theta=0.2$ rad, $\delta=r=1$ and $Ma_1=Ma_2=0.1$.}\label{f16}
\end{figure}

Figs.~\ref{f16}(a) and (b) exhibit the real and imaginary part of $\phi$, respectively, for different values of $m$. The mode shape variation for $\phi_r$ corresponding to the surface mode is negligible as compared to $\phi_i$. The velocity of fluid increases gradually towards the surface, however the velocity of fluid at interface dampens in the case of $m=2.5$ as compared to $m=0.5$ owing to the high viscous friction near the interface. This further reduces the velocity of fluid near the interface.

\begin{figure}[h!]
	\begin{center}
		\subfigure[]{\includegraphics*[width=7cm]{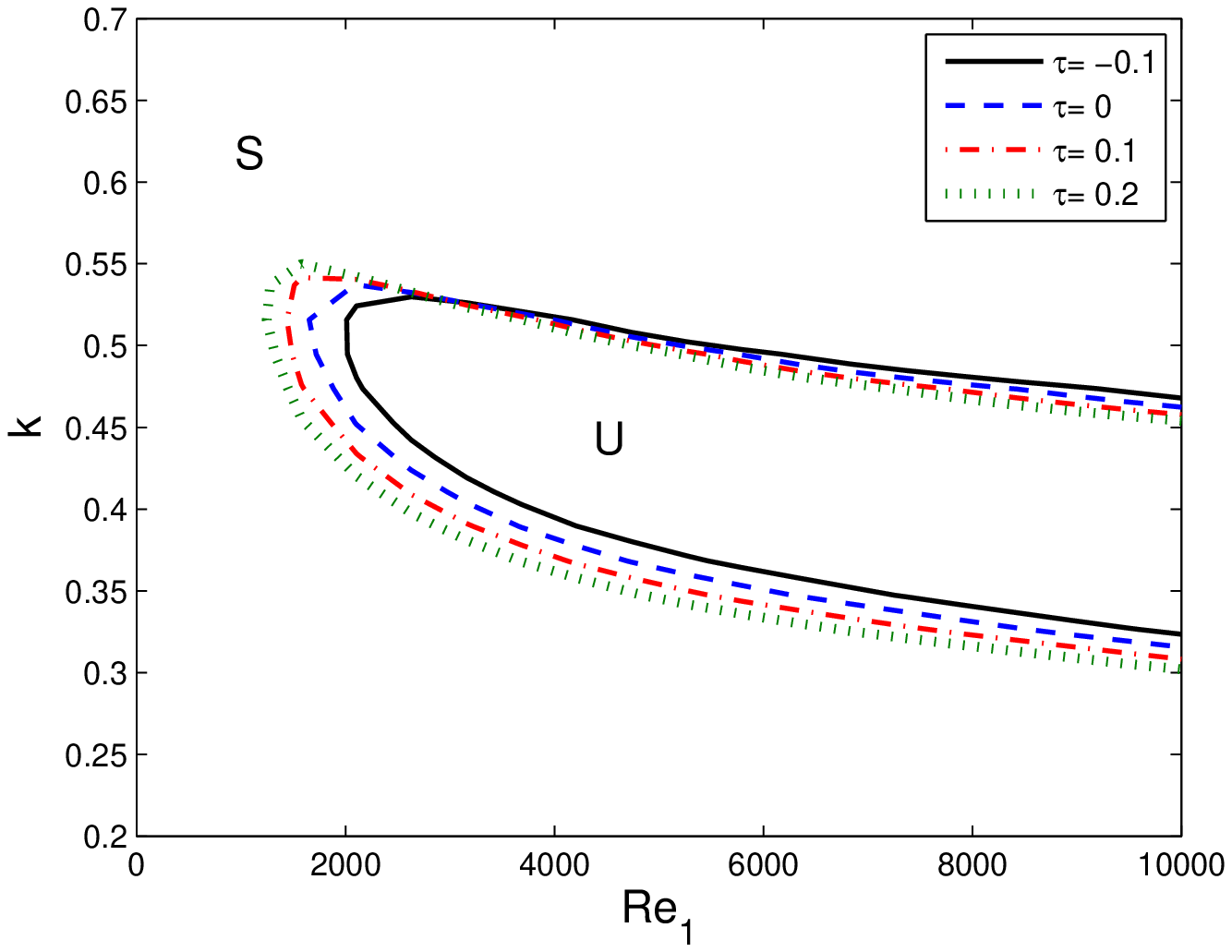}}
		\subfigure[]{\includegraphics*[width=7cm]{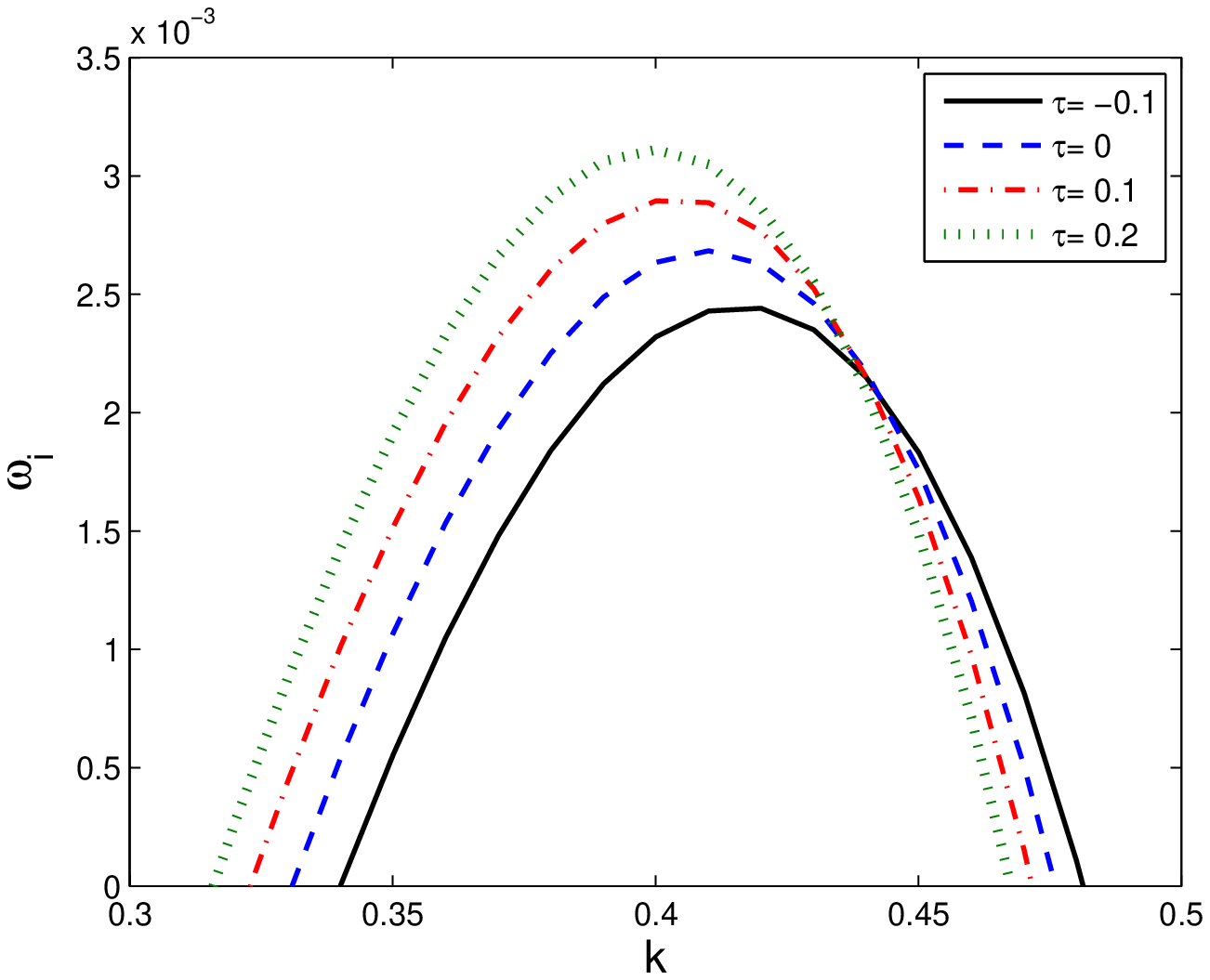}}
	\end{center}
	\caption{(a) Neutral stability curve corresponding to the varying external shear ($\tau$) and (b) its corresponding scaled growth rate ($\omega_i$) as a function of wavenumber ($k$) with $Re_1=8000$. The other constant flow parameters are $Pe_1=Pe_2=10000$, $Ca_1=Ca_2=1$, $\beta=0.04$, $\theta=0.2$ rad, $\delta=1$, $m=0.5$, $r=1.1$ and $Ma_1=Ma_2=0.1$.}\label{f17}
\end{figure}

\subsection{Effect of imposed shear for high Reynolds number}
	
In this subsection, the effect of external shear on the primary instability corresponding to the shear mode appearing at high Reynolds number is analyzed for the various flow parameters. Further, the analysis is made for sufficiently small inclination angle to compare the primary instability due to the surface and shear modes.

Fig.~\ref{f17}(a) depicts the neutral stability curves in the $(Re_1,k)$ plane showing the impact of the external shear force  ($\tau$). It is noticed that the onset of instability occurs early at lower critical $Re_1$ for higher value of $\tau$. The corresponding scaled growth rate as a function of wavenumber for various external shear is plotted in Fig.~\ref{f17}(a) at the Reynolds number $Re_1=8000$. It is worth to note that the external shear rate plays a dual role on the growth rate curves for a given value of $Re_1$. The temporal growth of the dominant mode raises for stronger imposed shear till a critical value of wave number ($k=0.44$) and a bifurcation is noticed thereafter ($k>0.44$).
This is validated with the result of marginal stability curves in Fig.~\ref{f17}(a) for the value of $Re_1=8000$. This dual variations of growth rate with respect to the wavenumber is due to the energy interaction between the surface/interfacial instabilities and the shear mode.

\begin{figure}[h!]
	\begin{center}
		{\includegraphics*[width=7cm]{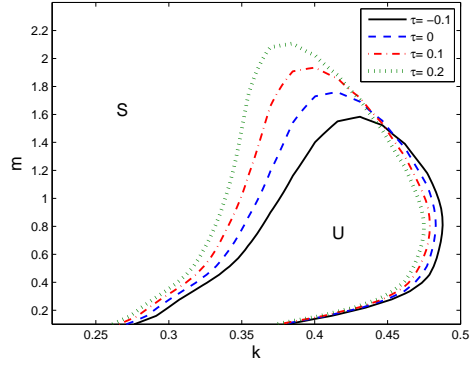}}
	\end{center}
	\caption{Marginal stability curve corresponding to the varying external shear ($\tau$) in the $(m,k)$ plane with $Re_1=8000$. The other constant flow parameters are $Pe_1=Pe_2=10000$, $Ca_1=Ca_2=1$, $\beta=0.04$, $\theta=0.2$ rad, $\delta=1$, $r=1.1$ and $Ma_1=Ma_2=0.1$.}\label{f18}
\end{figure}
\begin{figure}[h!]
	\begin{center}
		\subfigure[]{\includegraphics*[width=7cm]{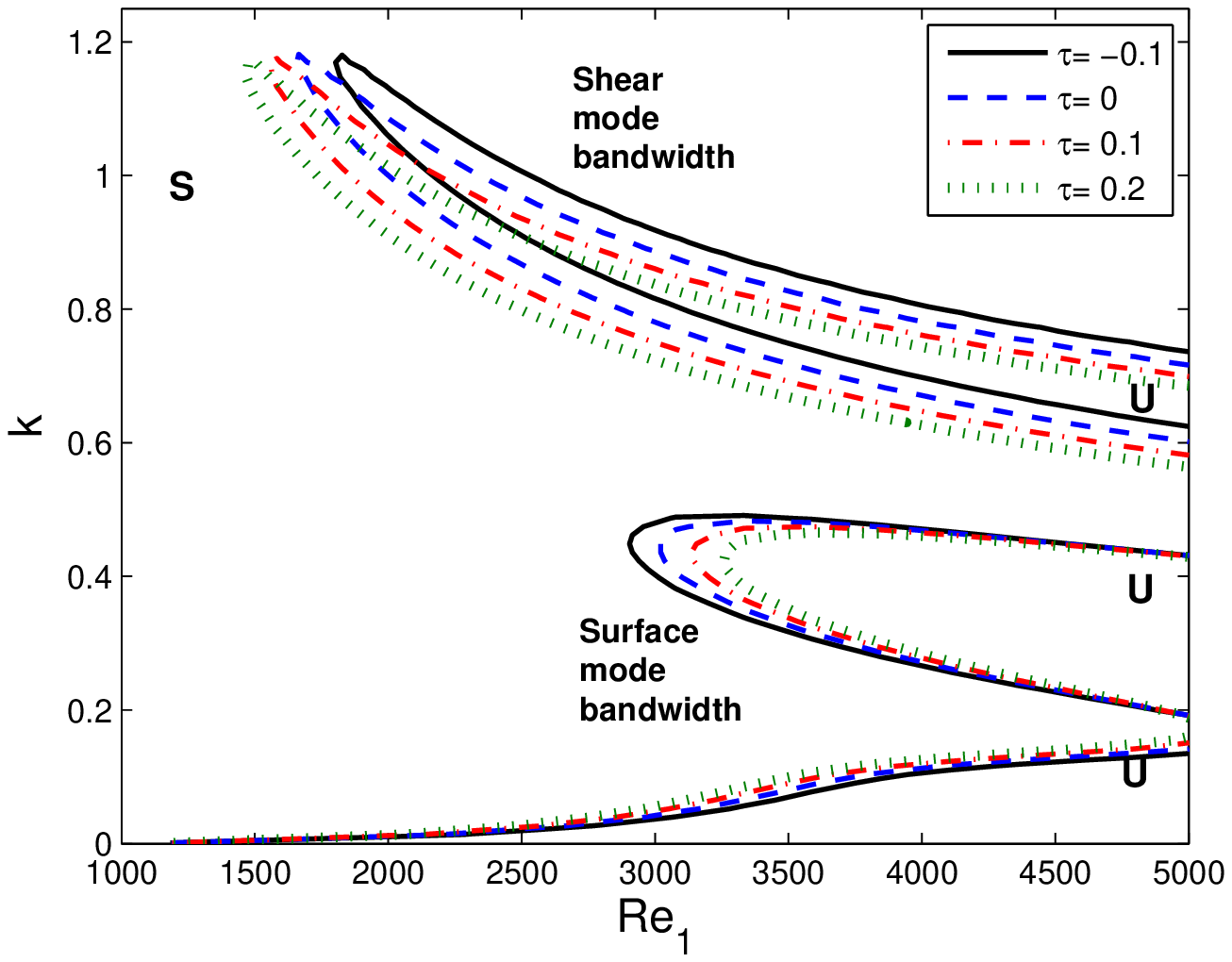}}
		\subfigure[]{\includegraphics*[width=7cm]{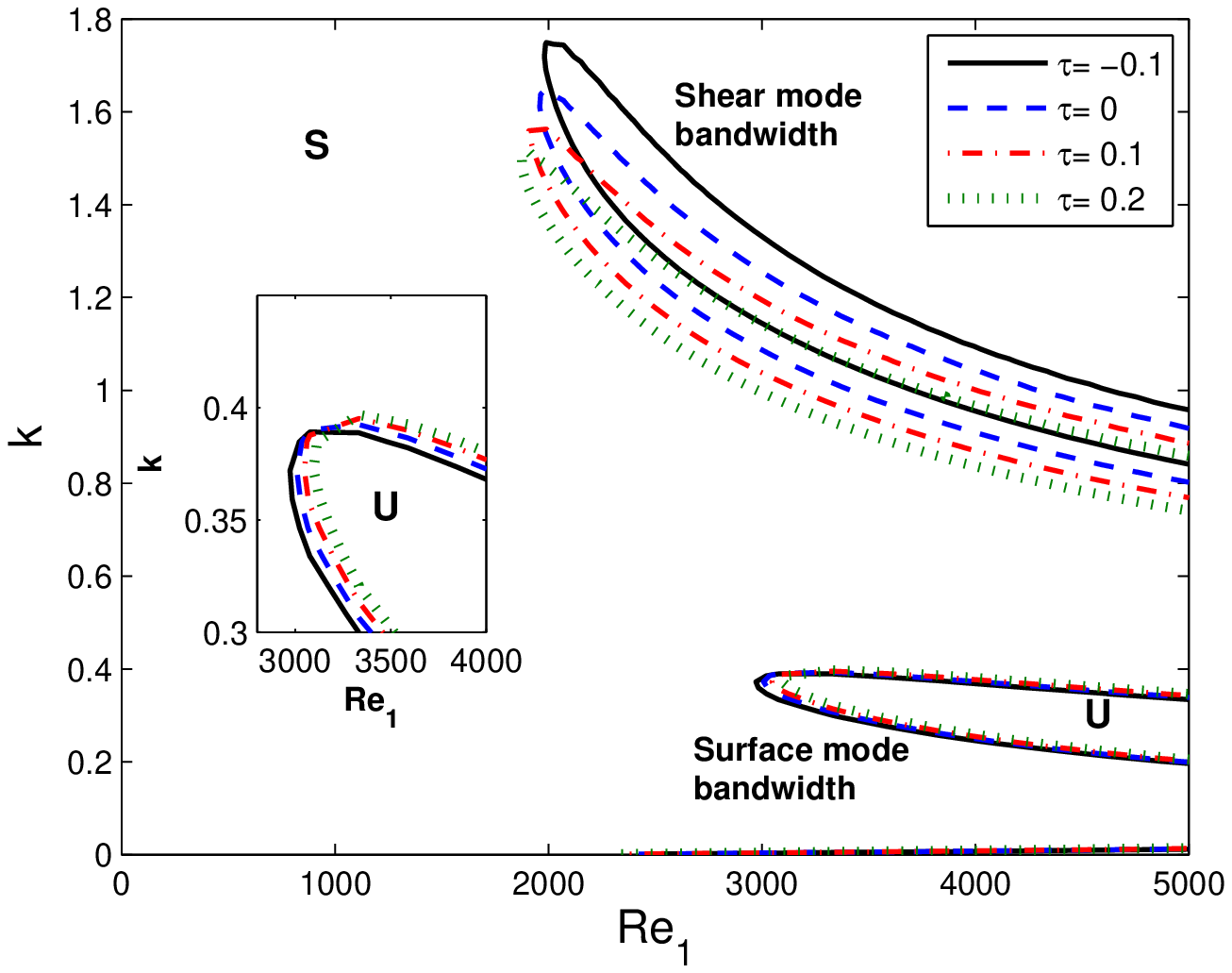}}
	\end{center}
	\caption{Marginal stability curve for the surface and shear modes in the $(Re_1,k)$ plane for (a) $\theta=0.6$ rad and (b) $\theta=0.5$ rad. The other constant flow parameters are $Pe_1=Pe_2=10000$, $Ca_1=Ca_2=1$, $\beta=0.04$, $\delta=r=1$, $m=0.5$, $r=1.1$ and $Ma_1=Ma_2=0.1$.}\label{f19}
\end{figure}

In Fig.~\ref{f18}, the marginal stability curves are portrayed including direct and opposite external shear effects in the $(m,k)$ plane. It is observed that the bandwidths are more significant for the smaller $m$ and the values of $m$ promotes more instability in the perturbed flow. On the other hand, for the larger values $m$ (i.e. $m>2.2$), there is a stable mode bandwidths enhancing the primary stability of the system. Further, the unstable shear mode bandwidth increases for increase in the value of $\tau$ and the observations in the $(m,k)$ are quit similar to that in the $(Re_1,k)$ plane (Fig.~\ref{f17}(a)).

The behavior of neutral stability curves with showing the role of external shear are plotted in the $(Re_1,k)$ plane for two different inclination angle $\theta=0.6'$ and $\theta=0.5'$ in Fig.~\ref{f19}(a) and (b), respectively. Further, the comparison study is made for both the shear and surface mode at very high Reynolds number range. It is observed from Fig.~\ref{f19}(a) that for $\theta=0.6'$ the unstable shear mode bandwidth increases with the superior imposed shear rate, whereas the unstable surface mode bandwidth sinks for increase in the external shear. At $\theta=0.6'$, the critical Reynolds number of surface mode is smaller than the shear mode and it implies that the surface mode dominates the shear mode at this situation. In Fig.~\ref{f19}(b) with $\theta=0.5'$, the unstable mode bandwidth fall short for higher range of external shear and the equivalent observation are noticed in the case of surface mode (shown in the inset of Fig.~\ref{f19}(b)). Further, the critical Reynolds number of shear mode is greater than the surface mode, which shows that the shear mode dominates over the surface mode at $\theta=0.5'$. Therefore, at the lower inclination angle shear mode becomes dominant over the surface mode as the role of the gravitational force to derive the surface mode is weaker for this configuration.
	
\section{Conclusions}\label{Conclusion}
	
The present study deals with the role of imposed external shear on the surfactant laden two-layer flow falling down a slippery pane, using the linear stability analysis. The governing equations and its associated boundary conditions are linearized applying the method of perturbation.  Further, by employing the normal mode solutions for the perturbed variables, the Orr--Sommerfeld equation corresponding to each layer is derived, and solved the coupled system utilizing the spectral collocation method. There exist different unstable eigenmodes, namely, surface mode, interface mode, and two surfactant modes assisting the surfactant transportation at the free surface and interface. In addition, at significantly high Reynolds numbers the shear modes are also recognized. The competitive effect of external shear rate on the unstable modes is analyzed by imposing the additional shear in the direction ($\tau > 0$) and counter direction ($\tau < 0$) of the flow. Moreover, based on the analysis and numerical results, the present study summarizes that: 

\begin{itemize}
\item At the moderate Reynolds numbers, both the surface and interfacial instabilities exist for certain range of wave numbers. The external shear has significant impact on the surface mode as well as interface mode. The behaviour of the interfacial instability varies depending on the viscosity ratio of the fluids. On considering the less viscous fluid in the lower layer ($m<1$), the shorter-wave and long-wave instabilities occur for the configurations $mr<1$ and $mr>1$, respectively, provided that the fluid in the lower layer is more dens ($r>1$). Irrespective of both configurations, the interfacial instabilities is enhanced for stronger external shear in the flow direction, whereas the extra shear in the opposite direction is favorable to stabilize the interface mode. Contrariwise, in the case of high viscous fluid in the lower layer with $r>1$, there exist a weaker unstable mode bandwidth in the shorter wave number region, in addition to the dominant unstable long-wave mode bandwidth. Quit interestingly, the higher imposed shear (with $\tau > 0$) shows a dual role, and triggers more shorter-wave instability and weaker long-wave instability. Further, the wall velocity slip has a stabilizing effect on the interface mode. The marginal stability curves confirm that the neutral condition for the interfacial instability is attained twice, resulting in the existence of the sub-critical Reynolds number.

\item A reverse impact of stronger external shear is noticed on surface mode instability. It is observed that the surface instabilities stabilizes with the increase in the external shear owing to the amplification of interfacial instabilities as a consequence of the momentum conservation. This is a fundamental finding and contrary to the role of imposed external shear on the surface mode for a surfactant laden single layer falling film. In the absence of external shear, analogous to the single fluid case, the wall velocity slip destabilizes the onset of surface mode instability but stabilizes it away from the critical Reynolds number. However, in the presence of strong external shear against the flow direction, the surface velocity reduces and resulting a overall stabilization of the surface mode by the slip parameter, which is again a new finding for the stratified two-layer falling film.

\item It is recognized that the unstable shear mode contributes to the primary instability of the flow for higher Reynolds number region. At a very small inclination angle the tangential component of the gravitational force is weak enough and consequently, the surface mode becomes less potent and the shear mode dominants. The results suggest for higher values of external shear rate, the shear mode bandwidth increases, however, the  opposite trends are observed in the case of surface mode. Further, on reducing the inclination angle, the shear mode befalls unstable early at lower critical Reynolds number than the surface mode. Moreover, to stabilize the shear mode one need to impose the superfluous shear along the reverse direction of the streamwise flow.

\end{itemize}  
\section*{Acknowledgment}
HB and SG gratefully acknowledge the financial support from SERB, Department of Science and Technology, Government of India through CRG project, Award No. CRG/2018/004521.\\

{\bf Data Availability:} The data that supports the findings of this study are available
within the article, highlighted in the related figure captions and corresponding discussions.

\bibliographystyle{unsrt}
\bibliography{references}

\end{document}